\documentclass[fleqn,10pt]{wlscirep}
\usepackage{graphicx,hyperref} 
\usepackage{nameref}

\usepackage{caption}
\usepackage{soul}
\usepackage{color}
\usepackage{stackengine}
\usepackage[position=t,singlelinecheck=off]{subcaption}
\usepackage{array}
\usepackage{multirow}
\usepackage{tabularx}
\usepackage{enumerate}
\usepackage[nameinlink]{cleveref} 
\usepackage{amsmath}
\usepackage{amssymb}
\usepackage{amsthm}
\usepackage{amsfonts}
\usepackage{amssymb}
\usepackage{rotating,pdflscape}
\usepackage{mathtools}
\usepackage{url}
\usepackage{bm}
\usepackage[utf8]{inputenc}
\usepackage[T1]{fontenc}
\usepackage[scaled]{helvet}%
\usepackage{lmodern}
\usepackage{microtype} 
\usepackage{xcolor}
\usepackage[normalem]{ulem}

\hypersetup{
  colorlinks   = true, 
  urlcolor     = black, 
  linkcolor    = blue, 
  citecolor   = blue 
}
\crefname{equation}{Eq.\!}{Eqs.\!}
\crefname{figure}{Fig.\!}{Figs.\!}

\def\Bmat{{\bf B}}
\def\Wmat{{\bf W}}

\def\Qcpmulti{Q^{\text{CP}} _{\gamma}}
\def\Qcp{Q^{\text{CP}}}
\def\ER{Erd\H{o}s-R\'{e}nyi\ }
\def\Exp{{\mathbb E}}
\newcommand\given[1][]{\:#1\vert\:}
\definecolor{myblue}{RGB}{33,150,243}
\definecolor{mygreen}{RGB}{76, 175, 80}
\definecolor{purple}{RGB}{170, 0, 255}

\renewcommand\footnotemark{}

\makeatletter
\renewcommand{\@seccntformat}[1]{}
\makeatother
\title{Multiscale core-periphery structure in a global liner shipping network}

\author[1,2,*]{Sadamori Kojaku\thanks{$\ ^*$These authors equally contributed to this work}}
\author[3,*]{Mengqiao Xu}
\author[3]{Haoxiang Xia}
\author[2,3,$\dagger$]{Naoki Masuda \thanks{$\ ^\dagger$naoki.masuda@bristol.ac.uk}}
\affil[1]{CREST, JST, Kawaguchi Center Building, 4-1-8, Honcho, Kawaguchi-shi, Saitama 332-0012, Japan}
\affil[2]{Department of Engineering Mathematics,
Merchant Venturers Building, University of Bristol,
Woodland Road, Clifton, Bristol, BS8 1UB, United Kingdom}
\affil[3]{
Faculty of Management and Economics, 
Dalian University of Technology, 
No.~2 Linggong Road, Ganjingzi District, Dalian City, Liaoning Province, 116024, China
}

\begin{abstract}
Maritime transport accounts for a majority of trades in volume, of which 70\% in value is carried by container ships that transit regular routes on fixed schedules in the ocean. 
In the present paper, we analyse a data set of global liner shipping as a network of ports. 
In particular, we construct the network of the ports as the one-mode projection of a bipartite network composed of ports and ship routes. 
Like other transportation networks, global liner shipping networks may have core-periphery structure, where a core and a periphery are groups of densely and sparsely interconnected nodes, respectively.
Core-periphery structure may have practical implications for understanding the robustness, efficiency and uneven development of international transportation systems.
We develop an algorithm to detect core-periphery pairs in a network, which allows one to find core and peripheral nodes on different scales and uses a configuration model that accounts for the fact that the network is obtained by the one-mode projection of a bipartite network.
We also found that most ports are core (as opposed to peripheral) ports and that ports in some countries in Europe, America and Asia belong to a global core-periphery pair across different scales, whereas ports in other countries do not.
\end{abstract}
\begin{document}

\flushbottom
\maketitle
%
%
\thispagestyle{empty}

\section{Introduction}
\label{sec:intro}
Transportation networks such as airways, railways and roadways underpin how the goods and people flow.
An understanding of the structure of transportation networks is crucial in finding bottleneck of transportation and vulnerable parts, contributing one to improve its efficiency and resilience \cite{Barthelemy2011}.  
Maritime transport is by far the most cost-effective way to move goods and raw materials across the globe. 
More than 80\% of global trade by volume is carried by ships and handled by seaports \cite{rev2017}. 
The most dominant type of global maritime transport in terms of seaborne trade value is the global liner shipping.
To date, container ships carry over 70\% value of the world trade \cite{rev2017}, making the global liner shipping network (GLSN) indispensable to the development of international trade and the world economy. 


Core-periphery (CP) structure is a meso-scale structure of networks that has been found in many networks including transportation networks such as airport networks \cite{Holme2005,Rossa2013,Kojaku2017,Kojaku2018a}, railway networks \cite{Rombach2017} and road networks \cite{Holme2005,Lee2014}. 
With CP structure based on edge density, a network is decomposed into a set of core nodes and that of peripheral nodes \cite{Borgatti2000,Boyd2006,Csermely2013,Lee2014,Tunc2015,Cucuringu2016,Kojaku2017,Rombach2017,Kojaku2018a}.
The nodes within the core are densely interconnected, those in the periphery are sparsely interconnected, and a node in the core and one in the periphery are connected with some probability depending on the assumption.
Previous studies suggested that transportation networks with CP structure would be robust against random failures (e.g., closure) of nodes \cite{Peixoto2012} and realise a competitive trade-off between the cost and profit \cite{Verma2016}.
Moreover, the existence of a core may contribute to the functional stability of networks \cite{Liu2011,Csermely2013}.

\begin{figure}
	\centering
	\begin{tabular}{c}
	\begin{minipage}[t]{0.3\hsize}
		\centering
		\includegraphics[width=\hsize]{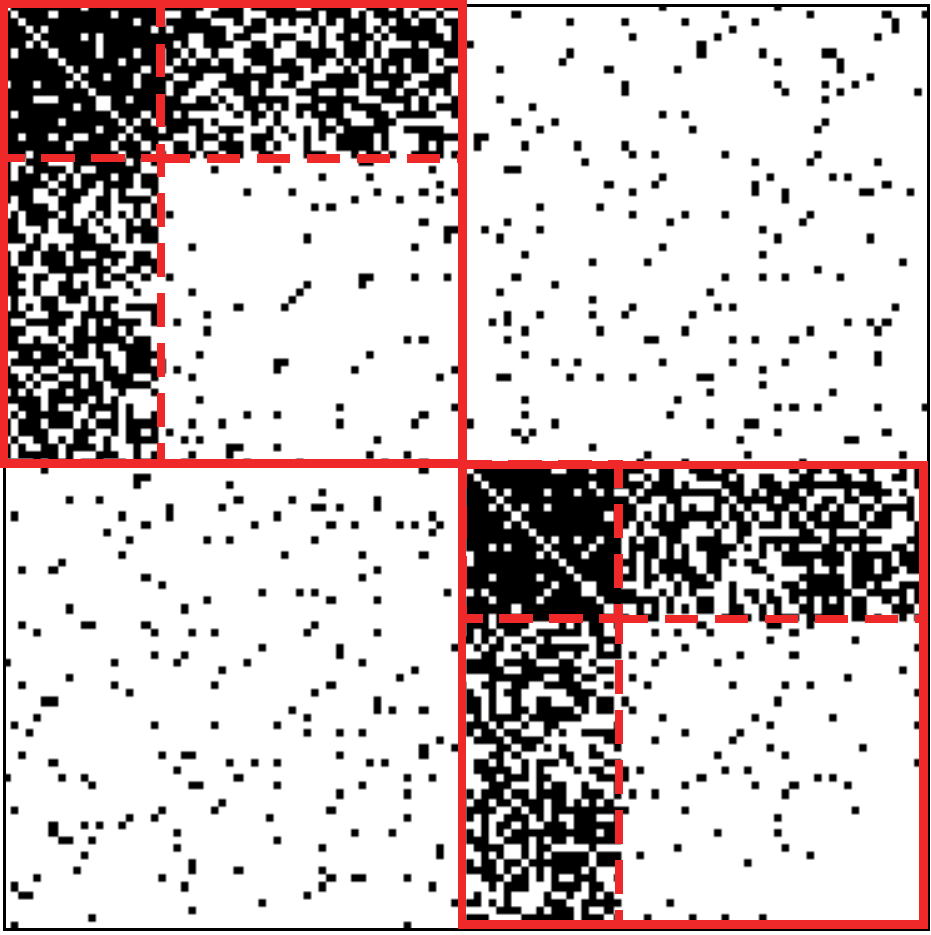}
	\end{minipage}
	\end{tabular}
	\caption{
	Adjacency matrix of a network with two CP pairs.
	The filled cell or empty cell indicates the presence or absence of an edge, respectively. 
	The solid line indicates the partition of nodes into the two CP pairs. 
	The dashed lines within each CP pair indicate the subpartition of nodes into the core and periphery.
	Each core block (top-left block in each CP pair) and periphery block (bottom right block in each CP pair) consist of 20 nodes and 40 nodes, respectively.
	The probability that each pair of nodes is adjacent by an edge is equal to 0.95 within each core block. 
	The same probability is equal to 0.8 between the core and periphery blocks within each CP pair.
	The same probability is equal to 0.05 within each periphery block and between different CP pairs.
	We draw edges according to these probabilities, independently for the different node pairs.
	}
	\label{fig:cp-example}
\end{figure}

The portrait of core-periphery dichotomy was postulated as a means to explain the uneven trade development and economic growth of nations in the process of globalization \cite{Krugman1995}. Maritime shipping serves as the primary transportation mode for international trade. As such, investigating the CP structure of the GLSN may help us to understand heterogeneous international trade among world regions and countries \cite{Mahutga2006,Garcia-Perez2016}. Specifically, there are many practical questions one can address by uncovering CP structure in maritime networks.
How can we plan shipping routes to improve the stability and economic efficiency of seaborne trade?
Which are the ports playing key roles in regional trade and those in international trades? 
How are ports integrated to global trade markets? Therefore, we analyse the CP structure in the GLSN. Crucially, we use the extension of our previous algorithm, Kojaku-Masuda (KM) algorithm \cite{Kojaku2017,Kojaku2018a}. The algorithm generally detects multiple CP pairs in networks (Fig.~\ref{fig:cp-example}), which many other algorithms do not.
We use this algorithm for two reasons. First, individual CP pairs are expected to correspond to either regional or global (or intermediate) groups of ports in each of which some ports may serve as core ports whereas the others may play a role of peripheral ports.
Second, in our previous studies \cite{Kojaku2017, Kojaku2018a}, the algorithm found CP pairs more accurately than other algorithms did in artificial networks with planted CP pairs.
We construct the GLSN from the empirical data on the liner shipping services operated by world's top 100 liner shipping companies in terms of fleet capacity (i.e., the twenty-foot equivalent unit capacity of the fleet). The data altogether account for over 92\% of the total fleet capacity in the world.

To reveal the CP structure in the GLSN, we extend our previous algorithm in the following three manners.
First, we adopt a null model that is compatible with the way we construct the GLSN from the data. 
Specifically, the original data set is regarded as a bipartite network composed of a layer of port nodes and a layer of shipping route nodes (Fig.~\ref{fig:one-mode-projection}).
Edges represent which ports belong to which shipping routes. 
Our null model discounts the effects induced by the one-mode projection of an originally bipartite network.
Second, our previous algorithms have a resolution limit, with which one can not find CP structure smaller than a threshold size  \cite{Kojaku2018a,Kojaku2018b}.
To circumvent this problem, we use a multiresolution method for community detection \cite{Reichardt2006,Heimo2008a} to extend the algorithm. 
Third, our previous algorithms provide different CP structures in the different runs of the same algorithm even if the initial condition is the same.
In the present study, we run the algorithm 100 times and look at the consensus of the results obtained from the different runs.

The present algorithm is applicable to networks constructed from a one-mode projection of bipartite networks. 
Examples of such networks include human disease networks \cite{Goh2007}, metabolic networks \cite{Guimera2005a} and mutualistic networks \cite{Padron2011}.
The Python code of the present algorithm is available on GitHub \cite{code}.


\section{Results}
\label{sec:results}
\subsection{Number of calling ports, number of serving routes, and node strength}

\begin{figure}
	\centering
	\begin{tabular}{cc}
	\includegraphics[width=\hsize]{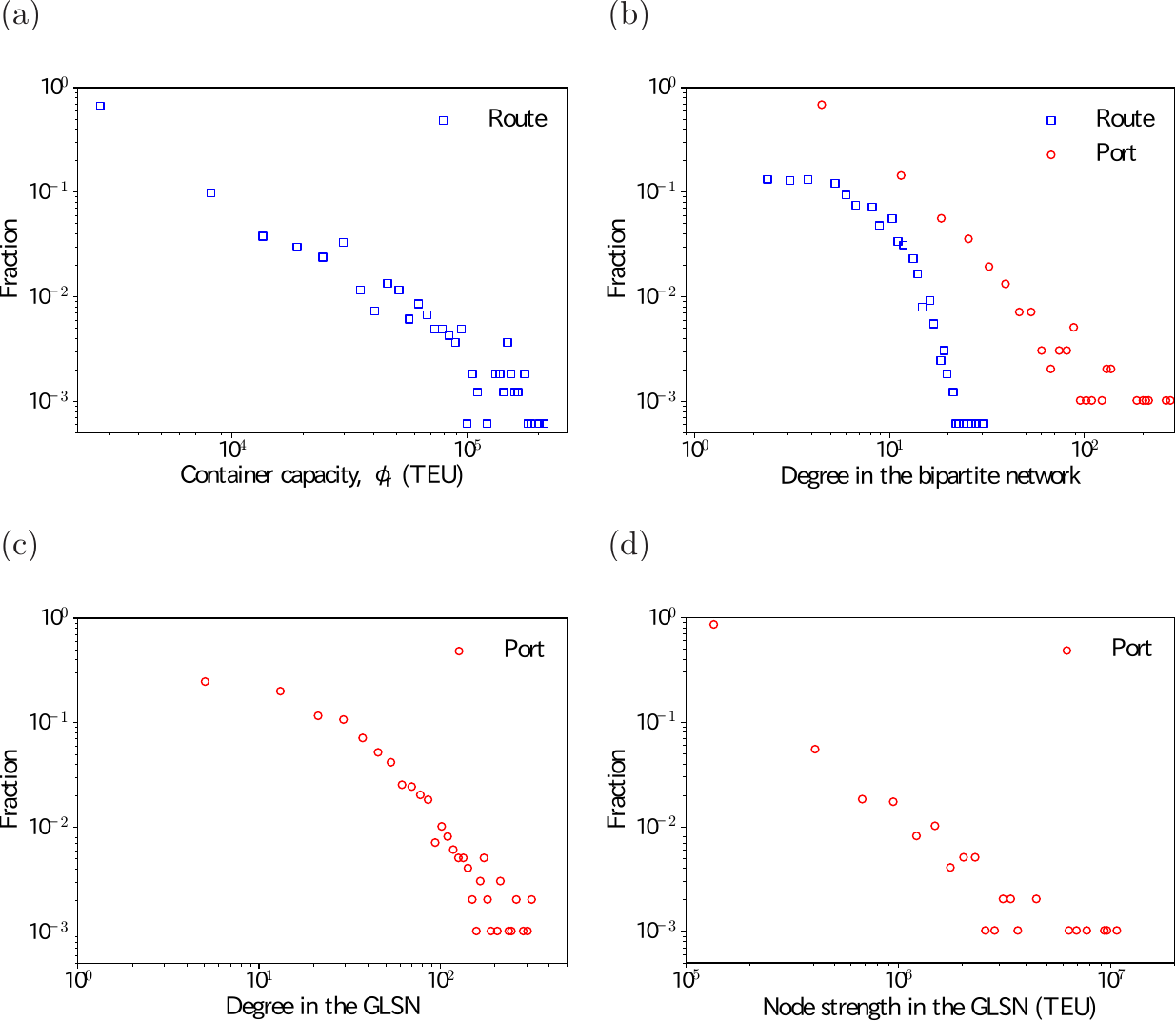}
	\end{tabular}
	\caption{
		Distributions of (a) container capacity $\phi_r$ of each route, (b) the node's degree in the bipartite network, 
		(c) the port's (unweighted) degree in the GLSN and 
		(d) the port's weighted degree (i.e., node strength) in the GLSN.
	}
	\label{fig:stat-bipartite}
\end{figure}

The distribution of the container capacity of a route (i.e., the sum of the maximum volume of containers that shipping companies deploy on the shipping route) is shown in Fig.~\ref{fig:stat-bipartite}(a).
The container capacity is heterogeneously distributed; a majority of the shipping routes has a capacity less than $10^2$, while 2\% of the routes has a capacity larger than $10^5$.
Degree $d_i ^{\text{port}}$ of ports in the bipartite network is also heterogeneously distributed (Fig.~\ref{fig:stat-bipartite}(b)).
A majority (56\%) of ports is shared by less than five routes, whereas 13 ports (1.3\%) including Shanghai and Singapore are shared by more than 100 routes.
Degree $d_r ^{\text{route}}$ of routes in the bipartite network is more homogeneously distributed than $d_i ^{\text{port}}$.
A majority (52\%) of routes contains less than five calling ports. 
The largest number of calling ports in a route is 31, which covers only $3.2\%$ of the $N=977$ ports.

The degree of each port in the GLSN is shown in Fig.~\ref{fig:stat-bipartite}(c).
A majority of ports (540 ports; 55\%) has a degree less than 25 in the GLSN, while 60 (6\%) ports have a degree larger than 100. 
We define node strength (i.e., weighted degree) of each port by the sum of the weight of edges attached to the port.
As is the case for the container capacity, node strength is heterogeneously distributed (Fig.~\ref{fig:stat-bipartite}(d)). 
Most ports (813 ports; 83\%) have a strength less than $2 \times 10^5$, while 51 ports (5\%) have a strength larger than $10^6$.

\subsection{Multiscale CP structure}
\label{sec:multi-cp}

We identify consensus CP pairs (we call them CP pairs for short in the following text) using the algorithm presented in the \nameref{sec:kmalgorithm} section.
The present algorithm is equipped with a resolution parameter $\gamma$, with which one can control the characteristic size of CP pairs to be detected.
Different $\gamma$ values may yield considerably different results. 
Therefore, we examine CP pairs across a range of $\gamma$, i.e., $\gamma \in \{0.01, 0.1, 0.2,0.3,\ldots,4\}$.

\begin{figure}
\centering
	\includegraphics[width=0.8\hsize]{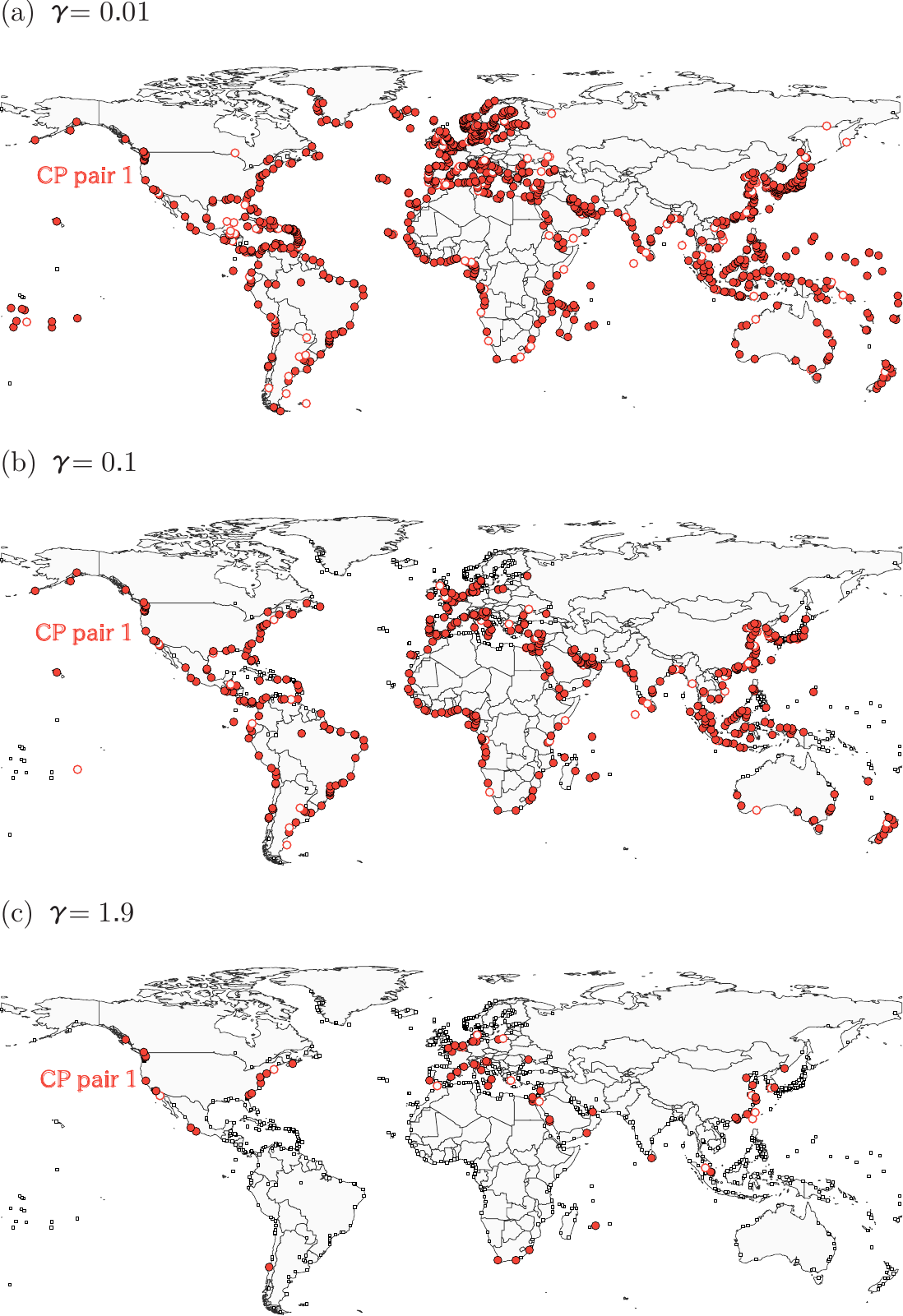}
\caption{
Consensus CP pairs in the GLSN. 
The resolution is equal to (a) $\gamma = 0.01$, (b) $\gamma = 0.1$, (c) $\gamma =1.9$.
The filled circles indicate the ports with a coreness value larger than 0.5. 
The open circles indicate the ports with a coreness value less than or equal to 0.5. 
The open squares indicate homeless ports.
}
\label{fig:cons_1}

\end{figure}

\begin{figure}
\centering
	\includegraphics[width=0.8\hsize]{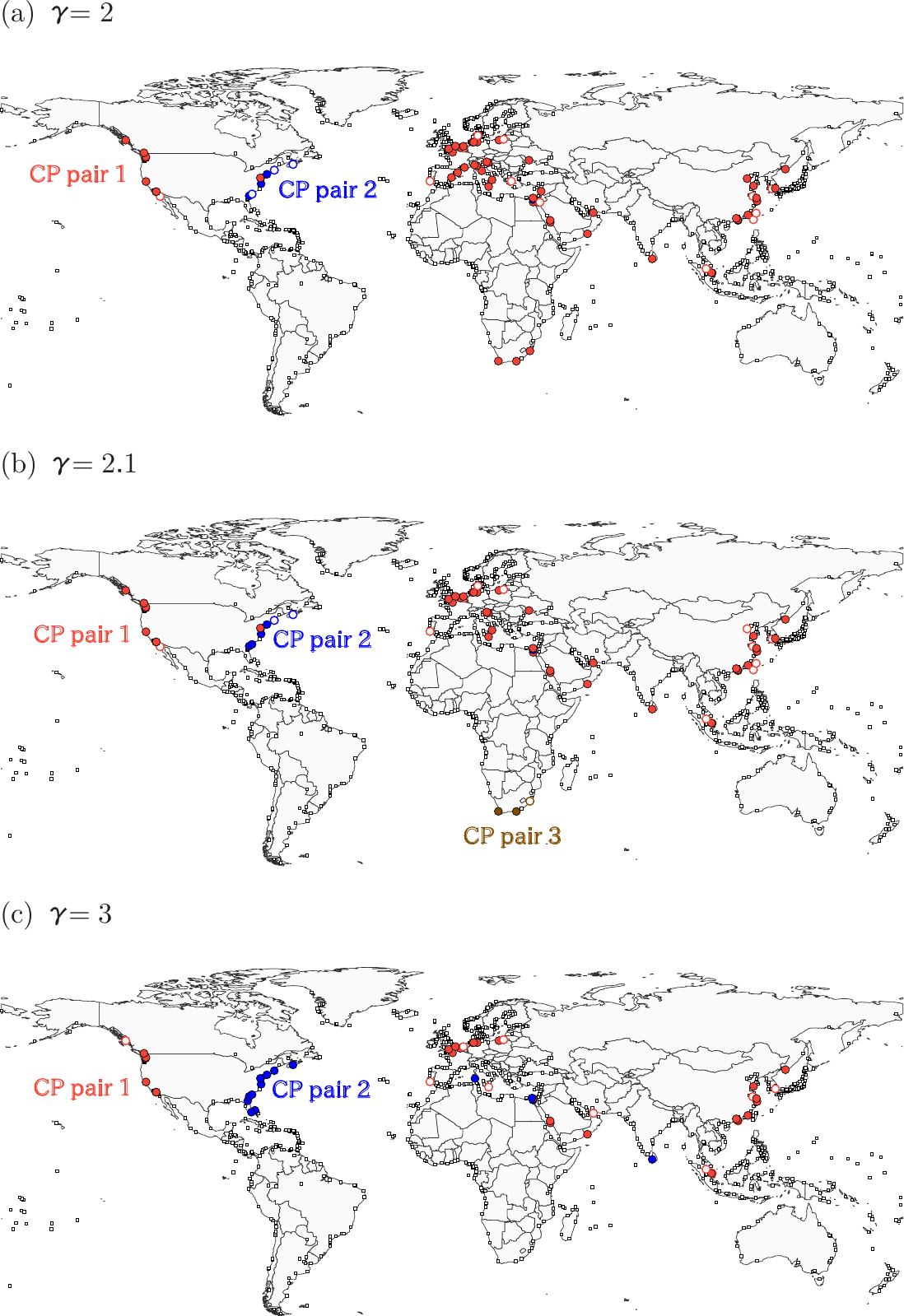}
\caption{
Consensus CP pairs in the GLSN. 
The resolution is equal to (a) $\gamma = 2$, (b) $\gamma = 2.1$, (c) $\gamma =3$.
}
\label{fig:cons_2}
\end{figure}

\begin{figure}
\centering
	\includegraphics[width=0.8\hsize]{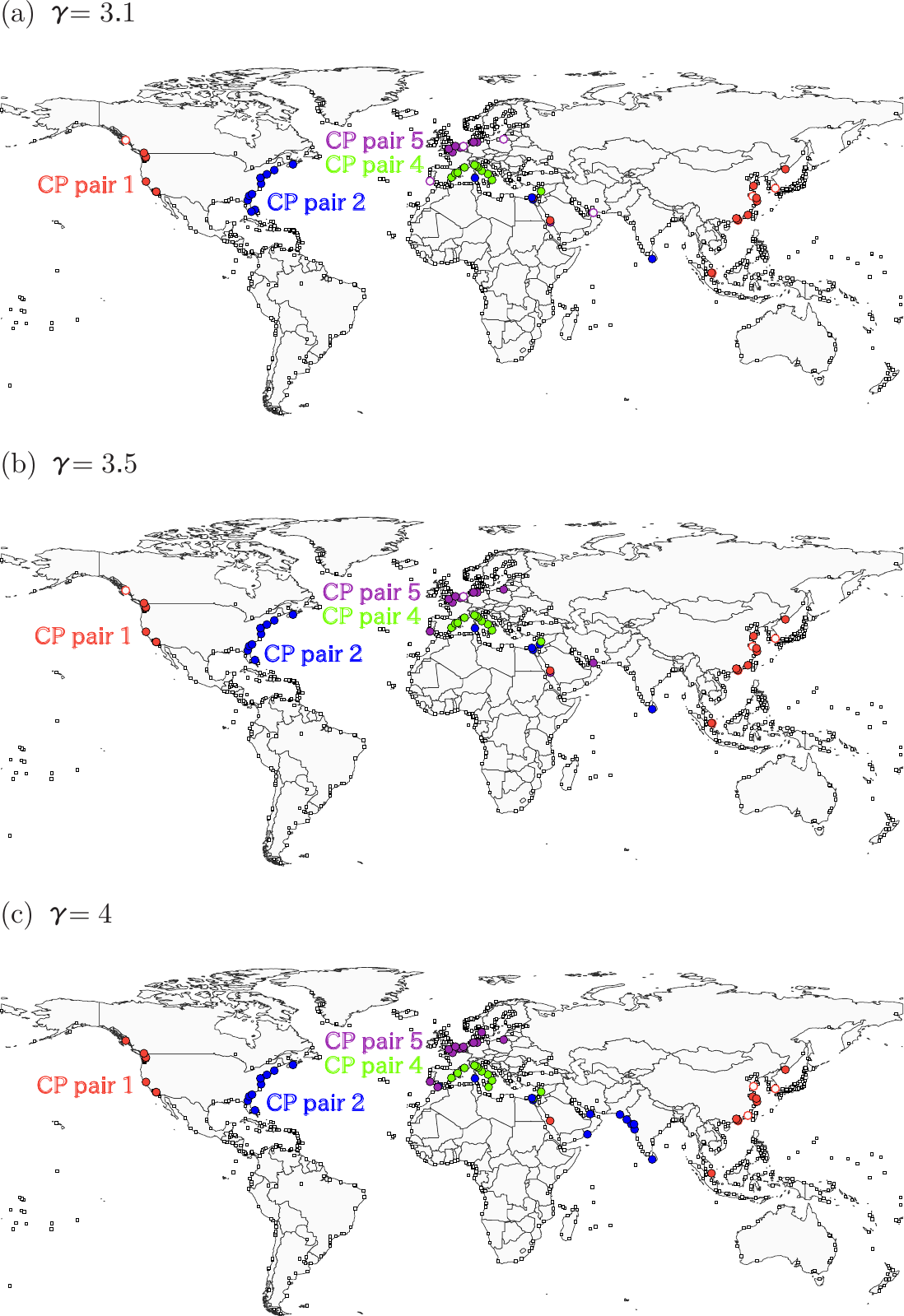}
\caption{
Consensus CP pairs in the GLSN. 
The resolution is equal to (a) $\gamma = 3.1$, (b) $\gamma = 3.5$ and (c) $\gamma =4$.
}
\label{fig:cons_3}
\end{figure}

We show the CP pairs detected at some $\gamma$ values in Figs.~\ref{fig:cons_1}--~\ref{fig:cons_3}.
There are at most five CP pairs. 
For $0.01 \leq \gamma \leq 1.9$, the algorithm identifies a unique CP pair containing ports in various geographical regions (Fig.~\ref{fig:cons_1}).
We refer to this CP pair as CP pair 1.
The number of ports in CP pair 1 decreases from 951 ports at $\gamma = 0.01$ to 76 ports at $\gamma = 1.9$.
At $\gamma =1.9$, the CP pair 1 contains many ports in China, the North Sea, the Mediterranean Sea and North America.
Few ports in Oceania, the South America, the West Africa and the East Africa belong to CP pair 1.

For $2 \leq \gamma \leq 3$, the algorithm identifies three CP pairs (Fig.~\ref{fig:cons_2}).
As is the case for $0.01 \leq \gamma \leq 1.9$, CP pair 1 contains the ports across many regions.
At $\gamma=2.0$, the algorithm identifies CP pair 2 that branches from CP pair 1 (Fig.~\ref{fig:cons_2}(a)).
CP pair 2 contains most ports in the East Coast of the US, a Canadian port (Halifax) and an Egyptian port (Suez).
At $\gamma = 2.1$, the algorithm identifies CP pair 3 located in the South Africa (Fig.~\ref{fig:cons_2}(a)).
CP pair 2 persists and enlarges in most cases as $\gamma$ increases. In contrast, CP pair 3 is absent for $\gamma \geq 2.2$ (Fig.~\ref{fig:cons_2}(b)).
 
For $3.1 \leq \gamma \leq 4$, the algorithm identifies four CP pairs. 
Each of CP pairs 1 and 2 spans different continents (Fig.~\ref{fig:cons_3}).
At $\gamma = 3.1$, CP pair 1 contains a majority of Chinese ports, the only port in Singapore and ports in the West Coast of the US.
CP pair 2 contains most ports in the East Coast of the US, two ports in the Mediterranean Sea, a port in Sri Lanka.
The other CP pairs 4 and 5 also branch from CP pair 1 and are composed of geographically close ports. 
In fact, CP pairs 4 and 5 mostly consist of the Mediterranean ports and North European ports, respectively.

\begin{figure}
\centering
\begin{tabular}{cc}
\begin{minipage}{0.6\hsize}
\centering
\includegraphics[width=\hsize]{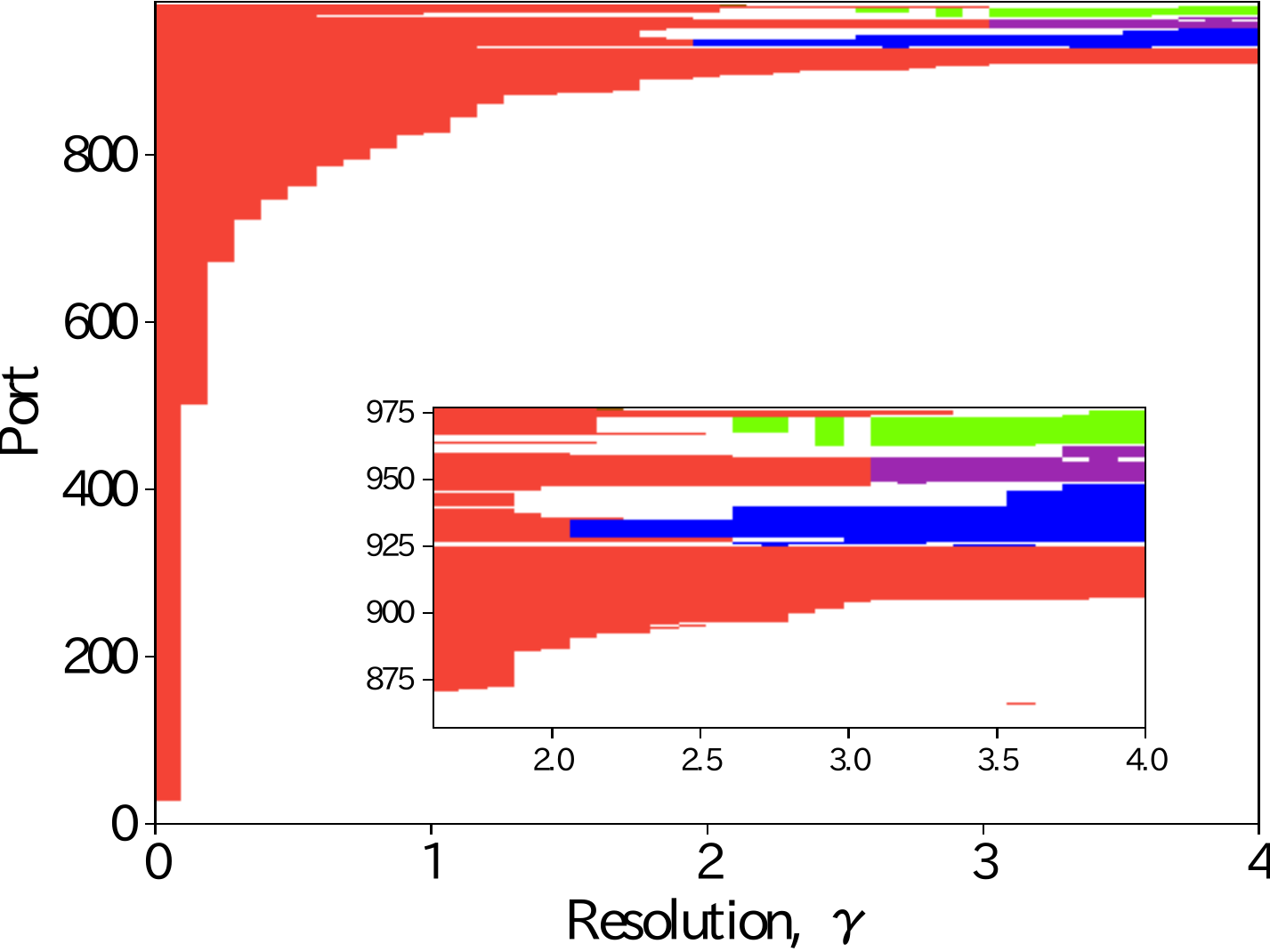}
\end{minipage}
& 
\begin{minipage}{0.2\hsize}
\centering
\includegraphics[width=0.7\hsize]{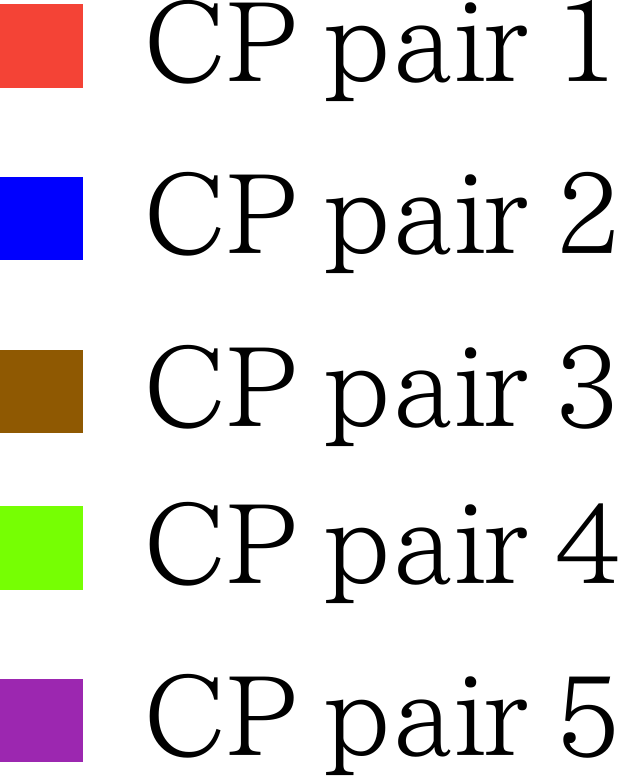}
\end{minipage}
\end{tabular}
\caption{
Membership of each port.
The colour at $(\gamma, i)$ indicates the index of the CP pair to which port $i$ belongs at resolution $\gamma$. 
The colour code is the same as that used in Figs.~\ref{fig:cons_1}--\ref{fig:cons_3}.
}
\label{fig:membership}
\end{figure}

The membership of each port at each $\gamma$ value is shown in Fig.~\ref{fig:membership}.
The number of ports in CP pair 1 decreases as $\gamma$ increases.
CP pairs 2, 4 and 5 detected for $2 \leq \gamma \leq 4$ are part of CP pair 1 detected for smaller $\gamma$ values.
CP pair 4 is absent for some $\gamma$ values for $2.6 \leq \gamma \leq 3$ but persists for $ 3.1 \leq \gamma \leq 4$.
As $\gamma$ increases, CP pairs 2, 4 and 5 largely expand by absorbing ports that belong to CP pair 1 at small $\gamma$ values.

\begin{figure}
\centering
\includegraphics[width=0.8\hsize]{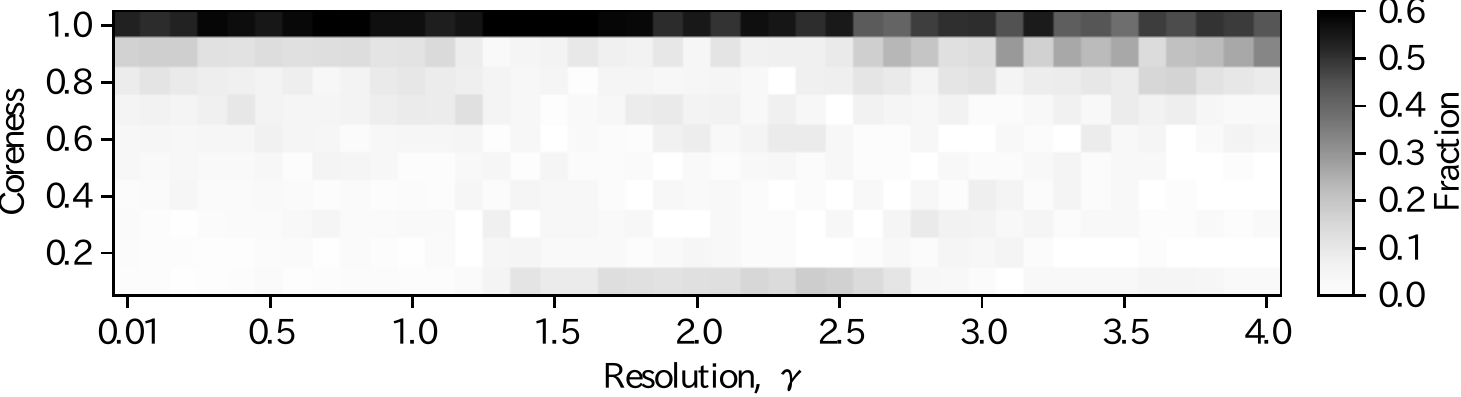}
\caption{
Distribution of coreness values of the ports in the consensus CP pairs. 
}
\label{fig:coreness}
\end{figure}

The distribution of the coreness values of ports in any CP pair is shown in Fig.~\ref{fig:coreness}.
For all $\gamma$ values, most ports have a coreness value larger than 0.9.
Therefore, the algorithm has classified most ports as core ports in most runs. 
If a CP pair only consists of core nodes, then the CP pair is a group of nodes that are densely interconnected with each other, which is equivalent to the usual notion of community.
Therefore, the current result indicates that the detected CP pairs are close to communities.
This property holds true for all $\gamma$ values that we have examined.

\subsection{Persistence of ports}

CP pair 1 considered across different resolutions (i.e., $\gamma$) has a nested relation.
In other words, CP pair 1 at resolution $\gamma$ contains CP pair 1 at all larger $\gamma$ values in a majority of cases.
This is the case for all but two ports when one varies $\gamma$ in the range $0.01\leq \gamma \leq 4$. 
Based on this observation, we define the persistence of a port as the smallest $\gamma$ value above which the port does not belong to CP pair 1 for the first time as one increases $\gamma$.
In other words, the persistence is the largest value of $\gamma$ such that the port belongs to CP pair 1 for all resolution values up to that $\gamma$ value. 
We note that the persistence is independent of $\gamma$.

\begin{figure}
\centering
\includegraphics[width=\hsize]{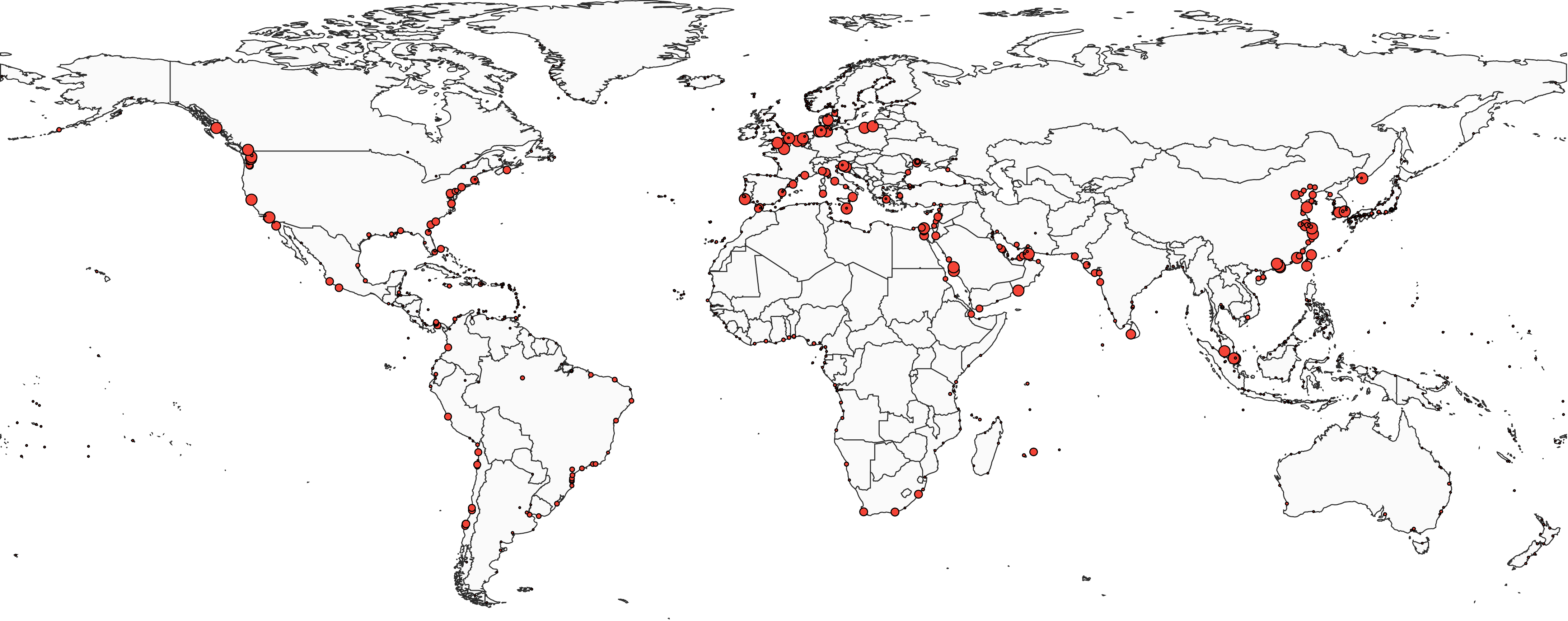}
\caption{
Persistence of each port, i.e., the largest resolution at which the port belongs to CP pair 1. 
The radius of the circle is proportional to the persistence of the port.  
}
\label{fig:persistence}
\end{figure}

The persistence of each port is represented by the size of the circle in Fig.~\ref{fig:persistence}. 
In the figure, only the ports belonging to CP pair 1 at $\gamma = 0.01$ are shown.
Highly persistent ports (e.g., persistence value larger than 3) are concentrated in China, the North Sea, the Mediterranean Sea, the Malay Peninsula, the Red Sea, and the West Coast of the US.
The two highly persistent ports in the Malay Peninsula, Singapore and Tanjung Pelepas, face the Strait of Malacca, which is an important shipping lane in the world \cite{Qu2012}. 
There are few highly persistent ports in the Caribbean Sea, Japan, Oceania, the East Coast of the South America, the East Africa and the West Africa.
Therefore, these regions may be relatively segregated from the main international shipping trade networks.

We show the ports with the persistence value larger than 2.8 in Table~\ref{ta:persistence}.
Highly persistent ports have a relatively large node strength (i.e., weighted degree).
More precisely, the persistence and node strength are positively correlated with the Spearman correlation coefficient being equal to 0.83.
We find that 497 ports (51\%) have a persistence value less than or equal to 0.1, while 64 ports (7\%) have a persistence value larger than 2.

\section{Discussion}
\label{sec:discussion}

We developed a multiscale algorithm to identify CP structure in a one-mode projection of bipartite networks, which intends to reveal multiscale CP pairs across different scales.
We applied the algorithm to a GLSN and revealed the inequality of regions in terms of the extent to which they are integrated into the global maritime transportation system. Specifically, our algorithm uncovered the following properties of the CP structure in the GLSN. 

First, at a coarse resolution, we detected a unique CP pair (CP pair 1) that mainly consists of ports in Asia, Europe and North America (Fig.~\ref{fig:cons_1}(c)). 
As major production and consumption centres on a global scale, these three regions have long been seen as dominating poles in global trade and container shipping activities \cite{Cesar2012}. 
Container shipping services that connect Asia and Europe, Asia and North America, and Europe and North America constitute the world's main East-West trading lanes, well-known as ``East-West Corridor'' in the maritime shipping industry \cite{Notteboom2008}.
Our result also provides some information on the integration of the economy in different regions into the global markets.
For instance, the ports in CP pair 1 are located in leading countries in trades (e.g., China, France, Germany, the United Kingdom and the United States) but not in Japan.
The absence of Japanese ports indicates that the integration of Japan into the global maritime transportation system may be insufficient, despite its status as the world's fourth-largest export economy in value. 
This situation might have a negative influence on the country's international trade development in the long run.

Second, for finer resolutions, the algorithm identified four small CP pairs that branch from CP pair 1 (Fig.~\ref{fig:cons_3}). 
These CP pairs involve main regional liner shipping markets of North Europe, Mediterranean, East Asia and North America, respectively. 
Two out of the four CP pairs, which are composed of major container ports in Northern Europe (CP pair 4) and the Mediterranean (CP pair 5), respectively, are geographically concentrated. In the liner shipping industry, they are highly developed conventional markets of intra-regional seaborne trade in Europe.
In contrast, the other two CP pairs extend across distinct geographical regions, corresponding to two inter-regional shipping routes in the West-East direction: North American East Coast-Mediterranean Sea-Indian Subcontinent shipping route via Suez Canal (CP pair 2) and North American West Coast-East Asia shipping route across the Pacific Ocean (CP pair 1). 
In particular, the dominance of China and the US in CP pair 1 is 
consistent with the high intensity of the bilateral trade between China and the US, the world's two largest countries in commodity trades \cite{UNCD}.

Third, the present algorithm classified a majority of ports in the GLSN as core ports as opposed to peripheral nodes (Fig.~\ref{fig:coreness}), as indicated by their high coreness values.
Although we do not know why this is the case, the result underlines the specificity of the GLSN. 
In fact, in worldwide airport networks, more than half of the airports were classified as peripheral nodes \cite{Kojaku2017, Kojaku2018a}.
This comparison indicates that the GLSN may be better regarded as a collection of communities, which is in agreement with the previous work reporting the community structure of global maritime shipping networks \cite{Kaluza2010}.
It should be noted that we found that CP pairs in the GLSN were similar to communities because we actually ran CP analysis.

Fourth, the persistence that we calculated for each port might be useful in evaluating the extent to which a port is integrated into the main international seaborne trade markets. The majority of the most persistent ports are regional load centres in the container shipping markets, i.e., world's leading container ports in terms of the yearly container throughput volume \cite{LloydList}.
Examples include East Asian ports of Busan, Guangzhou, Hong Kong, Ningbo-Zhoushan, Qingdao, Shanghai, and Shenzhen, Southeast Asian ports of Singapore and Tanjung Pelepas, North American West Coastal ports of Long Beach and Los Angeles, and European ports of Antwerp, Hamburg and Rotterdam.

Our study has the following limitations. 
First, we did not inform the edge weight by the actual container traffic between ports due to the commercial confidentiality. 
Instead, we used traffic capacity deployment data provided by shipping companies to approximate the actual traffic, assuming that the traffic capacity between any port pair on a same shipping service route was equal and bidirectional. 
Second, one-mode projection discards much information about the original bipartite network composed of the ports and routes. 
To mitigate this problem, one can use other one-mode projection methods that reflect some properties of the bipartite network to the projected networks \cite{Zhou2007,Gualdi2016,Saracco2017}.
Another approach is to study the original bipartite network without one-mode projection.
Third, we did not analyse another family of CP structure, i.e., transportation-based CP structure \cite{Holme2005,Csermely2013,Lee2014,Rombach2017}.
Transportation-based CP structure dictates that a core is a group of nodes that are frequently used in paths connecting nodes, e.g., nodes with high betweenness centrality.
Because GLSNs underlie maritime transportation, analysis of transportation-based CP structure may yield useful knowledge of the flow of cargo across the world.

\section{Methods}
\label{sec:methods}

\subsection{Data set}
\label{sec:dataset}

We use an empirical data set provided by Alphaliner \cite{Alphaliner}, which reports the statistics of $R=1,631$ major liner shipping service routes in the world for the year 2015. 
On each liner shipping service route (hereafter, shortened as service route), container ships call at a sequence of ports with a fixed service schedule. 
Cargo ships may call at ports for bunkering and maintenance, which are not directly associated with trade.
The present data set contains only the calling ports for cargo loading and unloading, ensuring a high relevance to world seaborne trade.
There are $N = 977$ ports in total.
We denote by $d^{\text{route}}_r$ the number of calling ports for route $r$.
Additionally, we denote by $d^{\text{port}}_i$ the number of routes that port $i$ serves.
The container capacity of route $r$, denoted by $\phi_r$, is given by the sum of the maximum volume of containers (counted in Twenty Equivalent Unit; TEU) deployed on shipping route $r$ by world shipping companies.

The data set does not contain the amount of containers transported between ports owing to the commercial confidentiality.
Therefore, we assume that the same amount of containers is transported between any pair of ports belonging to the same route. 
This procedure is equivalent to the following one-mode projection of the bipartite network.

\begin{figure}
	\centering
	\begin{tabular}{c}
	\begin{minipage}[t]{0.7\hsize}
		\centering
		\includegraphics[width=\hsize]{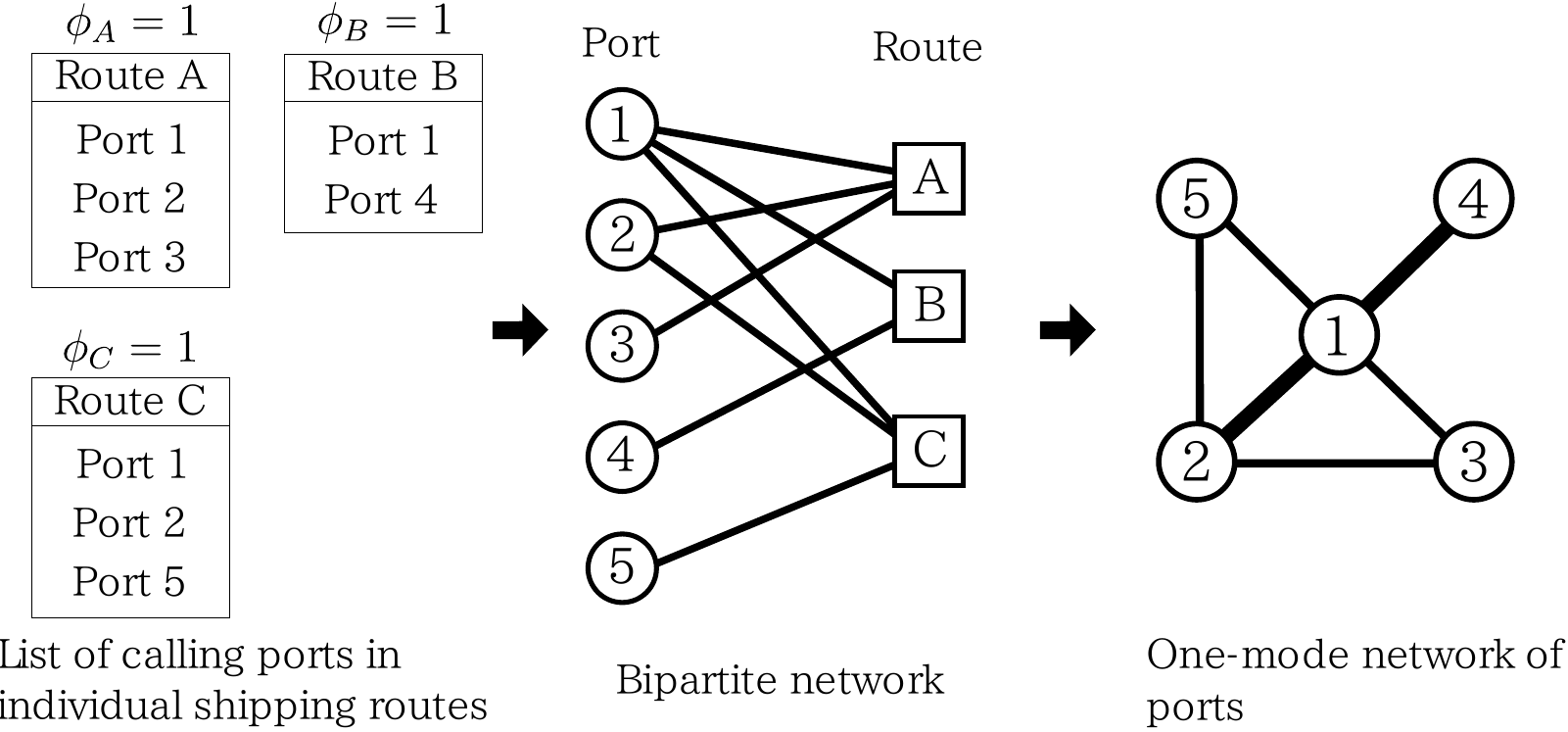}
	\end{minipage}
	\end{tabular}
	\caption{
		The construction of the GLSN.
		The width of edges in the one-mode network indicates the edge weight.
	}
	\label{fig:one-mode-projection}
\end{figure}

We represent the data as a bipartite network composed of ports and routes, where 
a port $i$ and a shipping route $r$ are adjacent if and only if port $i$ is a calling port of route $r$ (Fig.~\ref{fig:one-mode-projection}).
We denote by $\Bmat = (B_{ir})$ the $N \times R$ adjacency matrix of the bipartite network, where
$B_{ir} = 1$ or $B_{ir} = 0$ indicates that port $i$ and route $r$ are adjacent or not adjacent, respectively.

We construct the GLSN composed of ports by projecting the bipartite network to a one-mode network (Fig.~\ref{fig:one-mode-projection}). 
For example, in collaboration networks between academic authors, one connects all pairs of authors of a paper by an edge, resulting in a clique. 
Because a larger clique (i.e., a paper involving more authors) implies that the pairwise relationships between each pair of authors would be weaker, one often normalises the edge weight by dividing it by $d-1$  \cite{Guimera2007,Newman2010}, where $d$ is the number of authors of the paper.  
We apply the same method to the GLSN because the pairwise relationship between ports on a route would be relatively weak if the route involves many ports.
We assume that a route is worth a summed edge weight of unity for each port.
Then, we obtain
\begin{align}
    \label{eq:one-mode-projection}
    W _{ij} \equiv \left[1-\delta(i,j)\right] \sum_{r = 1}^ R \frac{\phi_r}{ d^{\text{route}}_r - 1 }B_{ir}B_{jr}, 
\end{align}
where $\delta(\cdot, \cdot)$ is Kronecker delta.
The sum of the weight of edges incident to each port (i.e., node strength) is equal to the sum of the container capacity deployed in all the individual service routes in which the port is involved.
This quantity is used for calculating the well-known country-level liner shipping connectivity index (LSCI) \cite{UNCTAD2}.
We note that the GLSN is a weighted network and does not contain self-loops (i.e., edges whose endpoints are the same node).

\subsection{Multiresolution algorithm}
\label{sec:kmalgorithm}

We regard a network as a collection of $C$ non-overlapping CP pairs (Fig.~\ref{fig:cp-example}). 
Each CP pair consists of one core block (i.e., group of nodes) and one periphery block.
By construction, there are many edges within each core block, whereas there are relatively few edges within each periphery block.
One may assume that there are many edges between the core and periphery blocks \cite{Borgatti2000,Cucuringu2016} or few edges \cite{Boyd2010,Craig2014}. 
We assume that there are many edges between the core and periphery blocks because we need to pair each periphery block with a particular core block. 

The present algorithm is an extension of our previous algorithm, which we call the KM algorithm \cite{Kojaku2017,Kojaku2018a}.
Therefore, we start by explaining the KM algorithm.
The algorithm identifies multiple CP pairs in networks, which many previous algorithms do not.
In the KM algorithm, we quantify the intensity of CP structure of a network by 
\begin{align}
    \label{eq:S}
    S \equiv \frac{1}{2\Omega}\sum_{i=1}^N \sum_{j=1}^{N} W_{ij}x_ix_j\delta(c_i,c_j) + \frac{1}{2\Omega}\sum_{i=1}^N \sum_{j=1}^{N} W_{ij}\left[ (1-x_i)x_j + x_i(x_j-1)\right]\delta(c_i,c_j),
\end{align}
where $c_i$ is the index of the CP pair to which node $i$ belongs, and $x_i = 1$ or $x_i = 0$ indicates that node $i$ is a core node or a peripheral node, respectively.
The first and second terms on the right-hand side of Eq.~\eqref{eq:S} are the fraction of the weight of edges confined within the core blocks and that connecting the core and periphery blocks within a CP pair, respectively.
Quantity $\Omega = \sum_{i=1} ^N \sum_{j=1}^{N} W_{ij}/2$ is the sum of the edge weight in the entire network, which normalises the value of $S$ between 0 and 1.
The KM algorithm seeks CP pairs by maximising  
\begin{align}
    \Qcp \equiv& S - \Exp[\tilde S] \nonumber \\
         =& \frac{1}{2\Omega}\sum_{i=1}^N \sum_{j=1}^{N} \left( W_{ij} -\Exp[\tilde W _{ij}]\right)(x_i + x_j - x_i x_j) \delta(c_i,c_j), \label{eq:qcpmulti-1}
\end{align}
where $\tilde S$ is the value of $S$ in a sample network generated from a null model. 
The adjacency matrix of the sampled network is denoted by $\tilde \Wmat = (\tilde W_{ij})$.
The expectation with respect to the null model is denoted by $\Exp[\cdot]$. 
We note that $\Qcp$ is equivalent to the modularity \cite{Newman2004,Reichardt2006} when all nodes are core nodes, i.e., $x_i = 1$ ($1 \leq i \leq N$).


This algorithm has a resolution limit \cite{Kojaku2018a}. 
In other words, CP pairs whose size is smaller than a threshold cannot be detected.
The modularity maximisation for finding communities in networks also shares this shortcoming  \cite{Fortunato2007}.
To discuss the CP structure at different resolutions, here we extend the algorithm \cite{Kojaku2018a,Kojaku2018b} using multiresolution methods \cite{Reichardt2006,Heimo2008a}.
In the new algorithm presented in this study, we seek CP pairs by maximising
\begin{align}
    \Qcpmulti \equiv 
         \frac{1}{2\Omega}\sum_{i=1}^N \sum_{j=1}^{N} \left( W_{ij} -\gamma \Exp[\tilde W _{ij}]\right)(x_i + x_j - x_i x_j) \delta(c_i,c_j), \label{eq:qcpmulti-1}
\end{align}
where $\gamma$ ($\gamma \geq 0$) is a resolution parameter that controls the effect of the null model term (i.e, $\Exp[\tilde W_{ij}]$).
The value of $\gamma$ affects the size of the CP pairs.
A detected CP pair is typically large if $\gamma$ is small. 
It should be noted that $\Qcpmulti$ is equivalent to $\Qcp$ when $\gamma = 1$.

The KM algorithm accepts various null models. We exploit this property to mitigate the artificial effect induced by the one-mode projection of bipartite networks such as the abundance of large cliques in the projected network.
In our previous algorithms \cite{Kojaku2017,Kojaku2018a}, we have adopted the \ER random graph \cite{erdHos1959random} or the configuration model \cite{Fosdick2016} as the null model.
With the configuration model, we rewire the edges by preserving the degree of each node; the \ER random graph does not preserve the degree of each node.
Here we use the configuration model as the null model because it is a standard null model in community detection \cite{Newman2004}, rich-club detection \cite{Colizza2006} and motif analysis \cite{Milo2002}.
However, applying the configuration model directly to the GLSN is problematic because the GLSN is obtained as the one-mode projection of a bipartite network (i.e., Eq.~\eqref{eq:one-mode-projection}). 
To circumvent this problem, we incorporate the effect of the one-mode projection into the configuration model, similar to a previous study on community detection \cite{Guimera2007}, as follows.

We generate a randomised bipartite network, whose adjacency matrix is denoted by $\tilde \Bmat = (\tilde B_{ir})$, using the configuration model. 
In other words, the randomised network preserves the degree of each node and the bipartiteness; otherwise, the network is uniformly randomly generated.
We allow multi-edges (i.e., multiple edges between the same pair of nodes) in the randomised bipartite networks for computational ease.
We carry out the one-mode projection of $\tilde \Bmat$ to obtain a randomised unipartite network, whose adjacency matrix is denoted by $\tilde \Wmat = (\tilde W_{ij})$.
The expected edge weight of the randomised unipartite network, $\Exp\left[\tilde W_{ij}\right]$, is given by
\begin{align}
    \label{eq:wij-1}
    \Exp[\tilde W_{ij}]  = \left[1-\delta(i,j) \right] \Exp\left[ \sum_{r = 1} ^R \frac{\phi_r}{ d_r ^{\text{route}} -1 } \tilde B_{ir} \tilde B_{jr} \right].
\end{align}
The randomised bipartite network (whose adjacency matrix is $\tilde \Bmat$) preserves the degree $d_r ^{\text{route}}$ of each route $r$.
Therefore, Eq.~\eqref{eq:wij-1} simplifies to
\begin{align}
    \label{eq:wij-2}
    \Exp[\tilde W_{ij}]  = \left[ 1-\delta(i,j) \right] \sum_{r = 1} ^R \frac{\phi_r}{ d_r ^{\text{route}} -1 } \Exp\left[\tilde B_{ir}  \tilde B_{jr} \right].
\end{align}
The term $\Exp\left[ \tilde B_{ir} \tilde B_{jr} \right]$ represents the probability that ports $i$ and $j$ are adjacent to route $r$ in the randomised bipartite network. 
With the configuration model, the probability that ports $i$ and $j$ are adjacent to route $r$ is equal to \cite{Guimera2007}
\begin{align}
    \label{eq:bir-bjr}
    \Exp\left[ \tilde B_{ir} \tilde B_{jr}\right] = d_i^{\text{port}} d_j^{\text{port}} \frac{d_r ^{\text{route}}(d_r ^{\text{route}}-1)}{M(M-1)},
\end{align}
where $M =\sum_{r^{\prime}=1} ^R d_{r^{\prime}} ^{\text{route}}$ is the number of edges in the randomised bipartite network. 
Substitution of Eq.~\eqref{eq:bir-bjr} into Eq.~\eqref{eq:wij-2} yields
\begin{align}
    \label{eq:wij-null}
    \Exp[\tilde W_{ij}]  = \left[ 1-\delta(i,j) \right] d_i^{\text{port}} d_j^{\text{port}} \sum_{r = 1} ^R \frac{\phi_rd_r ^{\text{route}}}{M(M-1)}.
\end{align}
By substituting Eq.~\eqref{eq:wij-null} into Eq.~\eqref{eq:qcpmulti-1}, we obtain the quality function 
\begin{align}
    \label{eq:qcpmulti-2}
    \Qcpmulti =\frac{1}{2\Omega}\sum_{i=1}^N \sum_{j=1}^{N} \left( W_{ij} -\gamma d_i^{\text{port}} d_j^{\text{port}} \sum_{r = 1} ^R \frac{\phi_r d_r ^{\text{route}}}{M(M-1)}\right)(x_i + x_j - x_i x_j) \delta(c_i,c_j). 
\end{align}

\subsection{Maximisation of $\Qcpmulti$}
\label{sec:algorithm}

We used a label switching heuristic to maximise $\Qcpmulti$ in our previous algorithms \cite{Kojaku2018a,Kojaku2018b}.
In our preliminary analysis, we found that the label switching heuristic in the present case detected multiple CP pairs in the GLSN for $\gamma=0$, whereas a single CP pair is natural anticipation in this case.
This result suggests that the label switching heuristic may return notably suboptimal results for various $\gamma$ values.  
Therefore, we implemented the following Louvain algorithm \cite{Blondel2008} to maximise the $\Qcpmulti$, which in fact yielded larger values of $\Qcpmulti$ than the label switching heuristic for all $\gamma$ values that we investigated.

\begin{figure}
	\centering
	\includegraphics[width=\hsize]{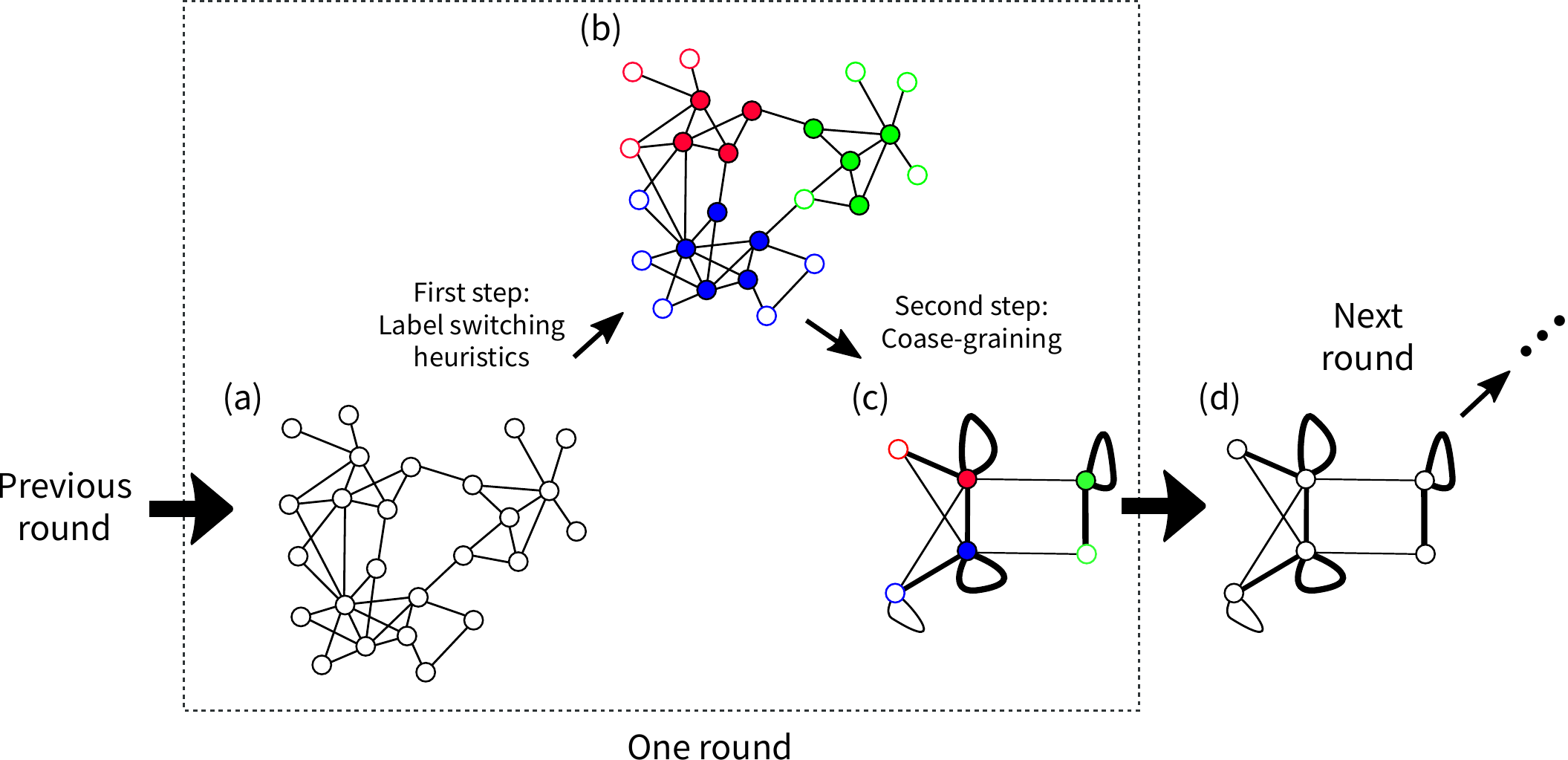}
	\caption{
		Schematic illustration of the variant of the Louvain algorithm.
		At the beginning of the current round, we have an input network of nodes (i.e., (a)).
		In the first step, we detect CP pairs in the input network using a label switching heuristic (i.e., (b)).
		In the second step, we construct a coarse-grained network by contracting the nodes in the input network having the same label into a super-node (i.e., (c)).
		Then, we perform the next round of which the input network is the coarse-grained network of the current round (i.e., (d)). 
		We iterate the rounds until the value of $\Qcpmulti$ stops increasing.
		(a) Input network for the current round.
		(b) CP pairs detected in the first step. 
		    The colour of each node indicates the CP pair to which the node belongs, i.e., $c_i$ $(1 \leq i \leq N)$.
		    The filled and blank circles indicate core and peripheral nodes, respectively, i.e., $x_i$. 
		(c) Coarse-grained network constructed in the second step.
		    The colour and openness of circles indicate the label $(c_i,x_i)$ of super-node $i$.
		    The thickness of the edge between super-nodes indicates the weight of the edge, i.e.,  
			the sum of the weight of the edges between a node in the input network belonging to one super-node and 
			a node in the input network belonging to the other super-node.  
		(d) The input network for the next round. 
	}
	\label{fig:alg-schematic}
\end{figure}

We iterate rounds, each of which consists of two steps (Fig.~\ref{fig:alg-schematic}).
In the first step, we identify CP pairs in a network using a label switching heuristic. 
In the second step, we coarse-grain the network by contracting the nodes belonging to the same CP pair detected in the first step into a super-node. 
(To avoid the confusion with the nodes in the original GLSN, here we use the term super-node to refer to a node in the coarse-grained network.) 
Then, we apply another round of the two steps to the coarse-grained network. 
We iterate the rounds of the two steps until the value of $\Qcpmulti$ stops increasing.
Then, we set the label of each node in the original network (i.e., $\Wmat$) to the label of the super-node to which it belongs in the final coarse-grained network.

The details of each step are as follows.
Let $\overline \Wmat$ be an $N' \times N'$ weighted adjacency matrix of the network in the beginning of the $r$th round, where 
$N'$ is the number of super-nodes in the beginning of the $r$th round.
We note that $\overline \Wmat = \Wmat$ and $N' = N$ in $r=1$.
In the first step of each round, we initialise the label of each super-node $i$ by $(c_i,x_i)= (i,1)$, where $1\leq i\leq N'$.
Then, we inspect each super-node in a random order.
For each inspected super-node $i$, we propose a new label $(c_i, x_i) = (c_j, 0)$, where super-node $j$ is a neighbour of super-node $i$ in the network specified by $\overline \Wmat$.
We also propose new label $(c_i, x_i) = (c_j, 1)$.
After carrying out this procedure for all neighbours of super-node $i$, we adopt the proposed label that yields the largest increment in $\Qcpmulti$.
If the largest increment in $\Qcpmulti$ is negative, then we do not change the label of super-node $i$.
The increment in $\Qcpmulti$ caused by changing the label of super-node $i$ from $(c,x)$ to $(c',x')$ is given by
\begin{align}
    \label{eq:dQ}
	&
		\frac{1}{\Omega} \Bigg[ 
				\overline W_{i,(c',1)} + x' \overline W_{i,(c',0)}
				-\overline W_{i,(c,1)} - x \overline W_{i,(c,0)}+(x' - x)\overline W_{ii}  \nonumber \\
				& - \gamma \overline d _i 
				\left( 
					\overline D_{(c',1)} + x'\overline D_{(c',0)}
					- \overline D_{(c,1)} -x\overline D_{(c,0)}
				\right)\left(\sum_{r = 1} ^R \frac{\phi_r d_r ^{\text{route}}}{M(M-1)}\right)
			\Bigg], 
\end{align}
where $\overline d_i$ is the sum of $d_{j} ^{\text{port}}$ values of the nodes belonging to super-node $i$, $\overline W_{i,(c,x)} = \sum_{j=1,j\neq i} ^{N'} \overline W_{ij} \delta(c,c_j)\delta(x,x_j)$ is the 
sum of the weight of the edges between super-node $i$ and other super-nodes with label $(c,x)$, and $\overline D_{(c,x)} = \sum_{j=1}  ^{N'}\overline d_j \delta(c,c_j)\delta(x,x_j)$ is the sum of $\overline d_i $ over the super-nodes with label $(c,x)$.
We note that  $\overline W_{ii} $ is the edge weight of the self-loop of super-node $i$.
If no label has changed in the process of inspecting the $N'$ super-nodes, then we proceed to the second step. 
Otherwise, we repeat to draw a new random order of the $N'$ super-nodes and inspect the $N'$ super-nodes for possible label switching, until no further increase in $\Qcpmulti$ occurs.

In the second step, we coarse-grain the network by contracting the super-nodes having the same label as a result of the first step into one super-node. 
In the new network, the edge weight between two super-nodes representing labels $(c,x)$ and $(c',x')$ is given by the sum of the weight of the edges between a super-node with label $(c,x)$ before the coarse-graining and a super-node with label $(c',x')$ before the coarse graining.
We note that the super-nodes may have self-loops (Fig.~\ref{fig:alg-schematic}). 

\subsection{Statistical test}
\label{sec:stat-test}

We examine the statistical significance of individual CP pairs using the so-called $(q,s)$--test \cite{Kojaku2018a,Kojaku2018b} that we previously proposed.
The $(q,s)$--test evaluates the significance of individual CP pairs.
For a CP pair in question, the $(q,s)$--test computes the quality of a CP pair composed of the same number of nodes in randomised networks.
Then, the $(q,s)$--test judges the CP pair in question as significant if its quality value is statistically larger than that of the CP pair of the same number of nodes in randomised networks.
The $(q,s)$--test requires a quality function $q$ for individual CP pairs.
We compute the quality of the CP pair $c$, denoted by $q_c$, by the contribution of the $c$th CP pair to $\Qcpmulti$, i.e., 
\begin{align}
    \label{eq:dq}
    q_c \equiv \frac{1}{2\Omega}\sum_{i=1}^N \sum_{j=1}^{N} \left( W_{ij} -\gamma d_i^{\text{port}} d_j^{\text{port}} \sum_{r = 1} ^R \frac{\phi_r d_r ^{\text{route}}}{M(M-1)}\right)(x_i + x_j - x_i x_j) \delta(c_i,c_j) \delta(c_i,c).
\end{align}
We note that the sum of $q_c$ over all CP pairs is equal to $\Qcpmulti$.

The value of $q_c$ would be positively correlated with the number $n_c$ of nodes in the $c$th CP pair \cite{Kojaku2018a}.
In other words, a large $q_c$ value may be caused by a large number of nodes in the CP pair.
To discount the effect of the correlation, the $(q,s)$--test assesses the significance of the $c$th CP pair using the conditional probability $P(\tilde q \geq q_c \given n_c)$ that the quality $\tilde q$ of a CP pair of the same size $n_c$ detected in a randomised network is larger than $q_c$.
If $P(\tilde q \geq q_c \given n_c)$ is smaller than a significance level $\alpha$ ($0 < \alpha \leq 1$), then one judges the CP pair in question to be significant. Otherwise, the CP pair is insignificant. 

In the $(q,s)$--test, one infers $P(\tilde q \geq q_c \given n_c)$ as follows.
First, we generate 500 randomised networks using the null model discussed in the \nameref{sec:kmalgorithm} section.
Second, we detect the CP pairs in the randomised networks using the present algorithm with the same resolution parameter used for finding the CP pair in question.
For each ${\overline c}$th detected CP pair in the 500 randomised networks, we compute the quality $\tilde q^{({\overline c})}$ and the number $\tilde n^{({\overline c})}$ of nodes in the CP pair.
Third, we infer a joint probability $P(\tilde q, \tilde n)$ using the Gaussian kernel density estimator \cite{Wand1993}, i.e., 
\begin{align}
    \label{eq:joint}
    P(\tilde q, \tilde n) &= \left.\sum_{{\overline{c}}=1} ^{\overline{C}} f\left( \frac{\tilde q - \tilde q^{({\overline{c}})}}{h\sigma_{\tilde q}}, \frac{\tilde n - \tilde n^{({\overline{c}})} }{h\sigma_{\tilde n}} \right) \middle/  \overline{C} \right.,
\end{align}
where $\overline C$ is the sum of the number of CP pairs detected in the 500 randomised networks,
and $\sigma_{\tilde q}$ and  $\sigma_{\tilde n}$ are the unbiased estimation of the standard deviation for $\{ \tilde q^{({\overline c})} \}$ and $\{ \tilde n ^{({\overline c})} \}$ ($1 \leq {\overline c} \leq {\overline C}$), respectively.  
Function $f(\cdot, \cdot)$ is the bivariate standard normal distribution given by  
\begin{align}
    \label{eq:bivariate}
    f(y_1,y_2) \equiv \frac{1}{2\pi \sqrt{1-\rho ^2}} \exp\left( -\frac{ y_1 ^2  - 2 \rho y_1 y_2 + y_2 ^2 }{2\left(1-\rho^2 \right)}  \right),
\end{align}
where $\rho$ is the Pearson correlation coefficient between $\{ \tilde q^{({\overline c})} \}$ and $\{ \tilde n ^{({\overline c})} \}$ ($1 \leq {\overline c} \leq {\overline C}$). 
Using Eq.~\eqref{eq:joint}, we obtain   
\begin{align}
    \label{eq:pval}
    P(\tilde q \geq q_c \given n_c ) &= 
			\frac{\int_{q_c} ^{\infty} P(\tilde q, n_c) {\rm d}\tilde q}{\int_{\infty} ^{\infty} P(\tilde q,n_c){\rm d}\tilde q} \nonumber \\	
		    & = 1 - \dfrac{
                    \displaystyle \sum\limits_{{\overline c}=1}^{{\overline C}}{ \exp\left( -\frac{\left(n_c - \tilde n ^{({\overline c})} \right)^2}{2\sigma_{\tilde n}^2h^2}\right)}
                    \Phi\left( \frac{\sigma_{\tilde n}\left(q_c -\tilde q^{({\overline c})}\right)  - \rho \sigma_{\tilde q}\left(n_c-\tilde n^{({\overline c})}\right)}{\sigma_{\tilde n}\sigma_{\tilde q}h\sqrt{1-\rho^2}}\right)
                    }{
                    \displaystyle \sum\limits_{{\overline c}=1}^{{\overline C}}{\exp\left( -\frac{\left(n_c -  \tilde n^{({\overline c})} \right)^2}{2\sigma_{\tilde n}^2h^2}\right)}
                    }, 
\end{align}
where $\Phi\left( y \right) = (2\pi)^{-1/2}\int^{y} _{-\infty} \exp(-u^2 /2){\rm d}u$ is the cumulative function of the standard normal distribution. 

We note that the Gaussian kernel estimator converges to any form of the probability distribution as the number of samples, ${\overline C}$, increases  \cite{Parzen1962}.  
Parameter $h$ is a free parameter that affects the speed of the convergence.
We use Scott's rule of thumb \cite{Scott2012}, i.e., $h={\overline C}^{\ -1/6}$.
We adopt the {\v{S}}id{\'{a}}k correction \cite{Sidak1967} to evade the multiple comparisons problem. 
In other words, we test each CP pair in the original network at a significance level of $\alpha = 1-(1-\alpha')^{1/C}$, where $\alpha'$ is the targeted significance. We set $\alpha'=0.05$.

\subsection{Consensus CP pairs}

Even one starts with the same initial condition, the present algorithm yields different significant CP structures in different runs due to the stochasticity of the algorithm. 
We address this issue by gathering the consensus of the results of different runs, which is regarded as a type of consensus clustering of data points \cite{Strehl2002,Topchy2005,Goder2008}.

To this end, we first run the present algorithm 100 times for a given value of $\gamma$.
(We show the results for 6 runs at each $\gamma$ value in the Supplementary Figures S1--S9).
Second, for each pair of ports $i$ and $j$, we compute the fraction of runs in which ports $i$ and $j$ belong to the same CP pair, which we denote by $P_{ij}$.
Third, we construct an undirected and unweighted network composed of the $N = 977$ ports, where two ports $i$ and $j$ are adjacent if and only if $P_{ij} \geq \theta$.
We set $\theta = 0.9$. 
Finally, we regard each connected component of the network as a consensus CP pair.
We refer to the ports that do not belong to any consensus CP pair as homeless ports. 
We define the coreness of each port $i$ in the consensus CP pair as the fraction of runs in which port $i$ is classified as core port.

\subsection{Matching CP pairs across resolutions}

Given consensus CP pairs calculated at different resolutions,
we match consensus CP pairs detected at two consecutive resolutions $\gamma$ and $\gamma'$ as follows.
For each consensus CP pair $c$ at resolution $\gamma$ and each consensus CP pair $c'$ at resolution $\gamma'$, we compute the similarity $\tau_{c,c'}$ between them using the Jaccard index, i.e., 
\begin{align}
	\tau_{c,c'} \equiv \frac{|V_{c} \cap V_{c'}|}{|V_{c} \cup V_{c'}|},
\end{align} 
where $V_{c}$ and $V_{c'}$ are the sets of ports in consensus CP pairs $c$ and $c'$, respectively.
We match $c$ and $c'$ if $\tau_{c,c'} > \max_{\overline c \neq c} \tau_{\overline c,c'}$ and $\tau_{c,c'} > \max_{\overline c \neq c'} \tau_{c,\overline c}$.
We note that some consensus CP pairs at resolution $\gamma$ may not be matched with any consensus CP pair at $\gamma'$ or vice versa. 
We did not find ties in the $\tau_{c,c'}$ value during the matching procedure.

As a result of the matching, we found seven consensus CP pairs across the resolution values. 
In fact, three of them (shown in green in Figs.~\ref{fig:cons_1}--\ref{fig:cons_3} and \ref{fig:membership}) are composed of almost the same set of nodes and reside in different ranges of $\gamma$ separated by gaps (therefore not contiguous in terms of the $\gamma$ value). 
Therefore, we regard these three consensus CP pairs as a single consensus CP pair.

\section{Competing interests}
The authors declare no competing interests.

\section{Author contributions}
M.~X. and N.~M. conceived and designed the research; 
M.~X. preprocessed the empirical data;
S.~K and N.~M. proposed the algorithm; S.~K performed the computational experiment; 
S.~K., M.~X., H.~X. and N.~M. wrote the paper.

\section{Acknowledgement}
M.~X. acknowledges the support provided through China Postdoctoral Science Foundation, Grant Number 2017M621141.
H. X. acknowledges the support provided through National Natural Science Foundation of China under Grant Numbers 71371040 and 71871042.
N.~M. acknowledges the support provided through JST, CREST, Grant Number JPMJCR1304.


\begin{thebibliography}{10}
\expandafter\ifx\csname url\endcsname\relax
  \def\url#1{\texttt{#1}}\fi
\expandafter\ifx\csname urlprefix\endcsname\relax\def\urlprefix{URL }\fi
\providecommand{\bibinfo}[2]{#2}
\providecommand{\eprint}[2][]{\url{#2}}

\bibitem{Barthelemy2011}
\bibinfo{author}{Barth{\'{e}}lemy, M.}
\newblock \bibinfo{title}{{Spatial networks}}.
\newblock \emph{\bibinfo{journal}{Phys. Rep.}} \textbf{\bibinfo{volume}{499}},
  \bibinfo{pages}{1--101} (\bibinfo{year}{2011}).

\bibitem{rev2017}
\bibinfo{note}{{Hoffmann J. et al., Review of maritime transport (United
  Nations Publications, 2017)}}.

\bibitem{Holme2005}
\bibinfo{author}{Holme, P.}
\newblock \bibinfo{title}{Core-periphery organization of complex networks}.
\newblock \emph{\bibinfo{journal}{Phys.~Rev.~E}} \textbf{\bibinfo{volume}{72}},
  \bibinfo{pages}{046111} (\bibinfo{year}{2005}).

\bibitem{Rossa2013}
\bibinfo{author}{Rossa, F.~D.}, \bibinfo{author}{Dercole, F.} \&
  \bibinfo{author}{Piccardi, C.}
\newblock \bibinfo{title}{{Profiling core-periphery network structure by random
  walkers}}.
\newblock \emph{\bibinfo{journal}{Sci.~Rep.}} \textbf{\bibinfo{volume}{3}},
  \bibinfo{pages}{1467} (\bibinfo{year}{2013}).

\bibitem{Kojaku2017}
\bibinfo{author}{Kojaku, S.} \& \bibinfo{author}{Masuda, N.}
\newblock \bibinfo{title}{Finding multiple core-periphery pairs in networks}.
\newblock \emph{\bibinfo{journal}{Phys. Rev. E}} \textbf{\bibinfo{volume}{96}},
  \bibinfo{pages}{052313} (\bibinfo{year}{2017}).

\bibitem{Kojaku2018a}
\bibinfo{author}{Kojaku, S.} \& \bibinfo{author}{Masuda, N.}
\newblock \bibinfo{title}{{Core-periphery structure requires something else in
  the network}}.
\newblock \emph{\bibinfo{journal}{New J. Phys.}} \textbf{\bibinfo{volume}{20}},
  \bibinfo{pages}{43012} (\bibinfo{year}{2018}).

\bibitem{Rombach2017}
\bibinfo{author}{Rombach, M.~P.}, \bibinfo{author}{Porter, M.~A.},
  \bibinfo{author}{Fowler, J.~H.} \& \bibinfo{author}{Mucha, P.~J.}
\newblock \bibinfo{title}{Core-periphery structure in networks (revisited)}.
\newblock \emph{\bibinfo{journal}{SIAM Rev.}} \textbf{\bibinfo{volume}{59}},
  \bibinfo{pages}{619--646} (\bibinfo{year}{2017}).

\bibitem{Lee2014}
\bibinfo{author}{Lee, S.~H.}, \bibinfo{author}{Cucuringu, M.} \&
  \bibinfo{author}{Porter, M.~A.}
\newblock \bibinfo{title}{Density-based and transport-based core-periphery
  structures in networks}.
\newblock \emph{\bibinfo{journal}{Phys.~Rev.~E}} \textbf{\bibinfo{volume}{89}},
  \bibinfo{pages}{032810} (\bibinfo{year}{2014}).

\bibitem{Borgatti2000}
\bibinfo{author}{Borgatti, S.~P.} \& \bibinfo{author}{Everett, M.~G.}
\newblock \bibinfo{title}{{Models of core/periphery structures}}.
\newblock \emph{\bibinfo{journal}{Soc.~Netw.}} \textbf{\bibinfo{volume}{21}},
  \bibinfo{pages}{375--395} (\bibinfo{year}{2000}).

\bibitem{Boyd2006}
\bibinfo{author}{Boyd, J.~P.}, \bibinfo{author}{Fitzgerald, W.~J.} \&
  \bibinfo{author}{Beck, R.~J.}
\newblock \bibinfo{title}{{Computing core/periphery structures and permutation
  tests for social relations data}}.
\newblock \emph{\bibinfo{journal}{Soc.~Netw.}} \textbf{\bibinfo{volume}{28}},
  \bibinfo{pages}{165--178} (\bibinfo{year}{2006}).

\bibitem{Csermely2013}
\bibinfo{author}{Csermely, P.}, \bibinfo{author}{London, A.},
  \bibinfo{author}{Wu, L.-Y.} \& \bibinfo{author}{Uzzi, B.}
\newblock \bibinfo{title}{Structure and dynamics of core/periphery networks}.
\newblock \emph{\bibinfo{journal}{J.~Comp.~Netw.}}
  \textbf{\bibinfo{volume}{1}}, \bibinfo{pages}{93} (\bibinfo{year}{2013}).

\bibitem{Tunc2015}
\bibinfo{author}{Tun{\c{c}}, B.} \& \bibinfo{author}{Verma, R.}
\newblock \bibinfo{title}{{Unifying inference of meso-scale structures in
  networks}}.
\newblock \emph{\bibinfo{journal}{PLOS ONE}} \textbf{\bibinfo{volume}{10}},
  \bibinfo{pages}{e0143133} (\bibinfo{year}{2015}).

\bibitem{Cucuringu2016}
\bibinfo{author}{Cucuringu, M.}, \bibinfo{author}{Rombach, P.},
  \bibinfo{author}{Lee, S.~H.} \& \bibinfo{author}{Porter, M.~A.}
\newblock \bibinfo{title}{{Detection of core–periphery structure in networks
  using spectral methods and geodesic paths}}.
\newblock \emph{\bibinfo{journal}{Eur.~J.~Appl.~Math.}}
  \textbf{\bibinfo{volume}{27}}, \bibinfo{pages}{846--887}
  (\bibinfo{year}{2016}).

\bibitem{Peixoto2012}
\bibinfo{author}{Peixoto, T.~P.} \& \bibinfo{author}{Bornholdt, S.}
\newblock \bibinfo{title}{{Evolution of robust network topologies: Emergence of
  central backbones}}.
\newblock \emph{\bibinfo{journal}{Phys. Rev. Lett.}}
  \textbf{\bibinfo{volume}{109}}, \bibinfo{pages}{118703}
  (\bibinfo{year}{2012}).

\bibitem{Verma2016}
\bibinfo{author}{Verma, T.}, \bibinfo{author}{Russmann, F.},
  \bibinfo{author}{Ara{{\'{u}}}jo, N.~A.~M.}, \bibinfo{author}{Nagler, J.} \&
  \bibinfo{author}{Herrmann, H.~J.}
\newblock \bibinfo{title}{{Emergence of core-peripheries in networks}}.
\newblock \emph{\bibinfo{journal}{Nat.~Commun.}} \textbf{\bibinfo{volume}{7}},
  \bibinfo{pages}{10441} (\bibinfo{year}{2016}).

\bibitem{Liu2011}
\bibinfo{author}{Liu, Y.-Y.}, \bibinfo{author}{Slotine, J.-J.} \&
  \bibinfo{author}{Barab{\'{a}}si, A.-L.}
\newblock \bibinfo{title}{{Controllability of complex networks}}.
\newblock \emph{\bibinfo{journal}{Nature}} \textbf{\bibinfo{volume}{473}},
  \bibinfo{pages}{167} (\bibinfo{year}{2011}).

\bibitem{Krugman1995}
\bibinfo{author}{Krugman, P.} \& \bibinfo{author}{Venables, A.~J.}
\newblock \bibinfo{title}{Globalization and the inequality of nations}.
\newblock \emph{\bibinfo{journal}{The Quarterly J. Econ.}}
  \textbf{\bibinfo{volume}{110}}, \bibinfo{pages}{857--880}
  (\bibinfo{year}{1995}).

\bibitem{Mahutga2006}
\bibinfo{author}{Mahutga, M.~C.}
\newblock \bibinfo{title}{{The persistence of structural inequality? A network
  analysis of international trade, 1965–2000}}.
\newblock \emph{\bibinfo{journal}{Soc. Forces}} \textbf{\bibinfo{volume}{84}},
  \bibinfo{pages}{1863--1889} (\bibinfo{year}{2006}).

\bibitem{Garcia-Perez2016}
\bibinfo{author}{Garc{\'{i}}a-P{\'{e}}rez, G.},
  \bibinfo{author}{Bogu{\~{n}}{\'{a}}, M.}, \bibinfo{author}{Allard, A.} \&
  \bibinfo{author}{Serrano, M.~{\'{A}}.}
\newblock \bibinfo{title}{{The hidden hyperbolic geometry of international
  trade: World Trade Atlas 1870--2013}}.
\newblock \emph{\bibinfo{journal}{Sci.~Rep.}} \textbf{\bibinfo{volume}{6}},
  \bibinfo{pages}{33441} (\bibinfo{year}{2016}).

\bibitem{Kojaku2018b}
\bibinfo{author}{Kojaku, S.} \& \bibinfo{author}{Masuda, N.}
\newblock \bibinfo{title}{{A generalised significance test for individual
  communities in networks}}.
\newblock \emph{\bibinfo{journal}{Sci. Rep.}} \textbf{\bibinfo{volume}{8}},
  \bibinfo{pages}{7351} (\bibinfo{year}{2018}).

\bibitem{Reichardt2006}
\bibinfo{author}{Reichardt, J.} \& \bibinfo{author}{Bornholdt, S.}
\newblock \bibinfo{title}{Statistical mechanics of community detection}.
\newblock \emph{\bibinfo{journal}{Phys.~Rev.~E}} \textbf{\bibinfo{volume}{74}},
  \bibinfo{pages}{016110} (\bibinfo{year}{2006}).

\bibitem{Heimo2008a}
\bibinfo{author}{Heimo, T.}, \bibinfo{author}{Kumpula, J.~M.},
  \bibinfo{author}{Kaski, K.} \& \bibinfo{author}{Saram{\"{a}}ki, J.}
\newblock \bibinfo{title}{{Detecting modules in dense weighted networks with
  the Potts method}}.
\newblock \emph{\bibinfo{journal}{J. Stat. Mechanics: Theory and Experiment}}
  \textbf{\bibinfo{volume}{2008}}, \bibinfo{pages}{P08007}
  (\bibinfo{year}{2008}).

\bibitem{Goh2007}
\bibinfo{author}{Goh, K.-I.} \emph{et~al.}
\newblock \bibinfo{title}{{The human disease network}}.
\newblock \emph{\bibinfo{journal}{Proc.~Natl.~Acad.~Sci.~USA}}
  \textbf{\bibinfo{volume}{104}}, \bibinfo{pages}{8685--8690}
  (\bibinfo{year}{2007}).

\bibitem{Guimera2005a}
\bibinfo{author}{Guimer{\`{a}}, R.} \& \bibinfo{author}{Amaral, L.~A.~N.}
\newblock \bibinfo{title}{{Functional cartography of complex metabolic
  networks}}.
\newblock \emph{\bibinfo{journal}{Nature}} \textbf{\bibinfo{volume}{433}},
  \bibinfo{pages}{895--900} (\bibinfo{year}{2005}).

\bibitem{Padron2011}
\bibinfo{author}{Padr{\'{o}}n, B.}, \bibinfo{author}{Nogales, M.} \&
  \bibinfo{author}{Traveset, A.}
\newblock \bibinfo{title}{{Alternative approaches of transforming bimodal into
  unimodal mutualistic networks. The usefulness of preserving weighted
  information}}.
\newblock \emph{\bibinfo{journal}{Basic and Appl. Ecol.}}
  \textbf{\bibinfo{volume}{12}}, \bibinfo{pages}{713--721}
  (\bibinfo{year}{2011}).

\bibitem{code}
\bibinfo{note}{{Kojaku, S. and Masuda, N. Python code of our algorithm.
  Available at https://github.com/skojaku/multiresolcp.}}

\bibitem{Qu2012}
\bibinfo{author}{Qu, X.} \& \bibinfo{author}{Meng, Q.}
\newblock \bibinfo{title}{{The economic importance of the Straits of Malacca
  and Singapore: An extreme-scenario analysis}}.
\newblock \emph{\bibinfo{journal}{Transp. Res. Part E: Logis. Transp. Rev.}}
  \textbf{\bibinfo{volume}{48}}, \bibinfo{pages}{258--265}
  (\bibinfo{year}{2012}).

\bibitem{Cesar2012}
\bibinfo{author}{C{\'{e}}sar, D.} \& \bibinfo{author}{Theo, N.}
\newblock \bibinfo{title}{{The worldwide maritime network of container
  shipping: spatial structure and regional dynamics}}.
\newblock \emph{\bibinfo{journal}{Global Netw.}} \textbf{\bibinfo{volume}{12}},
  \bibinfo{pages}{395--423} (\bibinfo{year}{2012}).

\bibitem{Notteboom2008}
\bibinfo{author}{Notteboom, T.} \& \bibinfo{author}{Rodrigue, J.-P.}
\newblock \bibinfo{title}{{Containerisation, box logistics and global supply
  chains: The integration of ports and liner shipping networks}}.
\newblock \emph{\bibinfo{journal}{Mari. Econom. {\&} Logis.}}
  \textbf{\bibinfo{volume}{10}}, \bibinfo{pages}{152--174}
  (\bibinfo{year}{2008}).

\bibitem{UNCD}
\bibinfo{note}{{United Nations Comtrade Database. UN Comtrade. Available at
  https://comtrade.un.org/data/ [Accessed: 22 Jul 2018]}}.

\bibitem{Kaluza2010}
\bibinfo{author}{Kaluza, P.}, \bibinfo{author}{K{\"{o}}lzsch, A.},
  \bibinfo{author}{Gastner, M.~T.} \& \bibinfo{author}{Blasius, B.}
\newblock \bibinfo{title}{{The complex network of global cargo ship
  movements}}.
\newblock \emph{\bibinfo{journal}{J. R. Soc. Interface}}
  \textbf{\bibinfo{volume}{7}}, \bibinfo{pages}{1093--1103}
  (\bibinfo{year}{2010}).

\bibitem{LloydList}
\bibinfo{note}{{Lloyd's List. Available at
  https://lloydslist.maritimeintelligence.informa.com/. [Accessed: 30 Jul
  2018]}}.

\bibitem{Zhou2007}
\bibinfo{author}{Zhou, T.}, \bibinfo{author}{Ren, J.}, \bibinfo{author}{Medo,
  M.} \& \bibinfo{author}{Zhang, Y.-C.}
\newblock \bibinfo{title}{{Bipartite network projection and personal
  recommendation}}.
\newblock \emph{\bibinfo{journal}{Phys. Rev. E}} \textbf{\bibinfo{volume}{76}},
  \bibinfo{pages}{046115} (\bibinfo{year}{2007}).

\bibitem{Gualdi2016}
\bibinfo{author}{Gualdi, S.}, \bibinfo{author}{Cimini, G.},
  \bibinfo{author}{Primicerio, K.}, \bibinfo{author}{{Di Clemente}, R.} \&
  \bibinfo{author}{Challet, D.}
\newblock \bibinfo{title}{{Statistically validated network of portfolio
  overlaps and systemic risk}}.
\newblock \emph{\bibinfo{journal}{Sci. Rep.}} \textbf{\bibinfo{volume}{6}},
  \bibinfo{pages}{39467} (\bibinfo{year}{2016}).

\bibitem{Saracco2017}
\bibinfo{author}{Saracco, F.} \emph{et~al.}
\newblock \bibinfo{title}{{Inferring monopartite projections of bipartite
  networks: an entropy-based approach}}.
\newblock \emph{\bibinfo{journal}{New J. Phys.}} \textbf{\bibinfo{volume}{19}},
  \bibinfo{pages}{53022} (\bibinfo{year}{2017}).

\bibitem{Alphaliner}
\bibinfo{note}{{Alphaliner. Available at https://www.alphaliner.com/ [Accessed:
  April 2015]}}.

\bibitem{Guimera2007}
\bibinfo{author}{Guimer{\`{a}}, R.}, \bibinfo{author}{Sales-Pardo, M.} \&
  \bibinfo{author}{Amaral, L. A.~N.}
\newblock \bibinfo{title}{{Module identification in bipartite and directed
  networks}}.
\newblock \emph{\bibinfo{journal}{Phys. Rev. E}} \textbf{\bibinfo{volume}{76}},
  \bibinfo{pages}{036102} (\bibinfo{year}{2007}).

\bibitem{Newman2010}
\bibinfo{author}{Newman, M.~E.~J.}
\newblock \emph{\bibinfo{title}{{Networks: An Introduction}}}
  (\bibinfo{publisher}{Oxford University Press}, \bibinfo{address}{Oxford},
  \bibinfo{year}{2010}).

\bibitem{UNCTAD2}
\bibinfo{note}{{United Nations Conference on Trade and Development. Available
  at http://unctadstat.unctad.org/wds/ReportFolders/reportFolders.aspx
  [Accessed: 10 July 2018]}}.

\bibitem{Boyd2010}
\bibinfo{author}{Boyd, J.~P.}, \bibinfo{author}{Fitzgerald, W.~J.},
  \bibinfo{author}{Mahutga, M.~C.} \& \bibinfo{author}{Smith, D.~A.}
\newblock \bibinfo{title}{{Computing continuous core/periphery structures for
  social relations data with MINRES/SVD}}.
\newblock \emph{\bibinfo{journal}{Soc.~Netw.}} \textbf{\bibinfo{volume}{32}},
  \bibinfo{pages}{125--137} (\bibinfo{year}{2010}).

\bibitem{Craig2014}
\bibinfo{author}{Craig, B.} \& \bibinfo{author}{von Peter, G.}
\newblock \bibinfo{title}{{Interbank tiering and money center banks}}.
\newblock \emph{\bibinfo{journal}{J.~Financ.~Intermed.}}
  \textbf{\bibinfo{volume}{23}}, \bibinfo{pages}{322--347}
  (\bibinfo{year}{2014}).

\bibitem{Newman2004}
\bibinfo{author}{Newman, M.~E.~J.} \& \bibinfo{author}{Girvan, M.}
\newblock \bibinfo{title}{Finding and evaluating community structure in
  networks}.
\newblock \emph{\bibinfo{journal}{Phys.~Rev.~E}} \textbf{\bibinfo{volume}{69}},
  \bibinfo{pages}{026113} (\bibinfo{year}{2004}).

\bibitem{Fortunato2007}
\bibinfo{author}{Fortunato, S.} \& \bibinfo{author}{Barth\'{e}lemy, M.}
\newblock \bibinfo{title}{{Resolution limit in community detection}}.
\newblock \emph{\bibinfo{journal}{Proc.~Natl.~Acad.~Sci.~USA}}
  \textbf{\bibinfo{volume}{104}}, \bibinfo{pages}{36--41}
  (\bibinfo{year}{2006}).

\bibitem{erdHos1959random}
\bibinfo{author}{Erd\H{o}s, P.} \& \bibinfo{author}{R{\'{e}}nyi, A.}
\newblock \bibinfo{title}{{On random graphs I}}.
\newblock \emph{\bibinfo{journal}{Publ.~Math.}} \textbf{\bibinfo{volume}{6}},
  \bibinfo{pages}{290--297} (\bibinfo{year}{1959}).

\bibitem{Fosdick2016}
\bibinfo{author}{Fosdick, B.}, \bibinfo{author}{Larremore, D.},
  \bibinfo{author}{Nishimura, J.} \& \bibinfo{author}{Ugander, J.}
\newblock \bibinfo{title}{{Configuring random graph models with fixed degree
  sequences}}.
\newblock \emph{\bibinfo{journal}{SIAM Rev.}} \textbf{\bibinfo{volume}{60}},
  \bibinfo{pages}{315--355} (\bibinfo{year}{2018}).

\bibitem{Colizza2006}
\bibinfo{author}{Colizza, V.}, \bibinfo{author}{Flammini, A.},
  \bibinfo{author}{Serrano, M.~A.} \& \bibinfo{author}{Vespignani, A.}
\newblock \bibinfo{title}{{Detecting rich-club ordering in complex networks}}.
\newblock \emph{\bibinfo{journal}{Nat.~Phys.}} \textbf{\bibinfo{volume}{2}},
  \bibinfo{pages}{110--115} (\bibinfo{year}{2006}).

\bibitem{Milo2002}
\bibinfo{author}{Milo, R.} \emph{et~al.}
\newblock \bibinfo{title}{{Network motifs: simple building blocks of complex
  networks}}.
\newblock \emph{\bibinfo{journal}{Science}} \textbf{\bibinfo{volume}{298}},
  \bibinfo{pages}{824--827} (\bibinfo{year}{2002}).

\bibitem{Blondel2008}
\bibinfo{author}{Blondel, V.~D.}, \bibinfo{author}{Guillaume, J.-L.},
  \bibinfo{author}{Lambiotte, R.} \& \bibinfo{author}{Lefebvre, E.}
\newblock \bibinfo{title}{{Fast unfolding of communities in large networks}}.
\newblock \emph{\bibinfo{journal}{J.~Stat.~Mech.}}
  \textbf{\bibinfo{volume}{2008}}, \bibinfo{pages}{P10008}
  (\bibinfo{year}{2008}).

\bibitem{Wand1993}
\bibinfo{author}{Wand, M.~P.} \& \bibinfo{author}{Jones, M.~C.}
\newblock \bibinfo{title}{{Comparison of smoothing parameterizations in
  bivariate kernel density estimation}}.
\newblock \emph{\bibinfo{journal}{J.~American Stat.~Assoc.}}
  \textbf{\bibinfo{volume}{88}}, \bibinfo{pages}{520--528}
  (\bibinfo{year}{1993}).

\bibitem{Parzen1962}
\bibinfo{author}{Parzen, E.}
\newblock \bibinfo{title}{{On estimation of a probability density function and
  mode}}.
\newblock \emph{\bibinfo{journal}{Annal. Math. Stat.}}
  \textbf{\bibinfo{volume}{33}}, \bibinfo{pages}{1065--1076}
  (\bibinfo{year}{1962}).

\bibitem{Scott2012}
\bibinfo{author}{Scott, D.~W.}
\newblock \bibinfo{title}{{Multivariate density estimation and visualization}}.
\newblock In \emph{\bibinfo{booktitle}{Handbook of Computational Statistics}},
  \bibinfo{pages}{549--569} (\bibinfo{publisher}{Springer, Berlin},
  \bibinfo{year}{2012}).

\bibitem{Sidak1967}
\bibinfo{author}{{\v{S}}id{\'{a}}k, Z.}
\newblock \bibinfo{title}{{Rectangular confidence regions for the means of
  multivariate normal distributions}}.
\newblock \emph{\bibinfo{journal}{J.~Am.~Stat.~Assoc.}}
  \textbf{\bibinfo{volume}{62}}, \bibinfo{pages}{626--633}
  (\bibinfo{year}{1967}).

\bibitem{Strehl2002}
\bibinfo{author}{Strehl, A.} \& \bibinfo{author}{Ghosh, J.}
\newblock \bibinfo{title}{{Cluster ensembles – A knowledge reuse framework
  for combining multiple partitions}}.
\newblock \emph{\bibinfo{journal}{J. Machi. Learning Res.}}
  \textbf{\bibinfo{volume}{3}}, \bibinfo{pages}{583--617}
  (\bibinfo{year}{2002}).

\bibitem{Topchy2005}
\bibinfo{author}{Topchy, A.}, \bibinfo{author}{Jain, A.~K.} \&
  \bibinfo{author}{Punch, W.}
\newblock \bibinfo{title}{{Clustering ensembles: models of consensus and weak
  partitions}}.
\newblock \emph{\bibinfo{journal}{IEEE Trans. Patt. Anal. and Machi. Intel.}}
  \textbf{\bibinfo{volume}{27}}, \bibinfo{pages}{1866--1881}
  (\bibinfo{year}{2005}).

\bibitem{Goder2008}
\bibinfo{author}{Goder, A.} \& \bibinfo{author}{Filkov, V.}
\newblock \bibinfo{title}{{Consensus clustering algorithms: Comparison and
  refinement}}.
\newblock In \emph{\bibinfo{booktitle}{Proc. Meeting on Alg. Eng. {\&}
  Experiments}}, \bibinfo{pages}{109--117} (\bibinfo{publisher}{Society for
  Industrial and Applied Mathematics}, \bibinfo{address}{Philadelphia, PA,
  USA}, \bibinfo{year}{2008}).

\end{thebibliography}

\clearpage

\clearpage

\clearpage

\clearpage

\clearpage

\clearpage

\clearpage

\clearpage

\begin{table}
\caption{Ports with the largest persistence values.}
\label{ta:persistence}
\centering
\begin{tabular}{c}
\begin{minipage}{0.5\hsize}
\begin{tabular}{llrc}
Name & Country & Strength & Persistence\\ \hline \hline
Shanghai & China & $10,831,283$ & $4.0$\\
Shenzhen & China & $9,646,887$ & $4.0$\\
Ningbo-Zhoushan & China & $9,415,002$ & $4.0$\\
Hong Kong & China & $6,880,967$ & $4.0$\\
Busan & South Korea & $6,309,888$ & $4.0$\\
Qingdao & China & $4,369,577$ & $4.0$\\
Xiamen & China & $3,416,973$ & $4.0$\\
Tanjung Pelepas & Malaysia & $3,116,464$ & $4.0$\\
Guangzhou & China & $2,361,131$ & $4.0$\\
Oakland & US & $1,953,242$ & $4.0$\\
Los Angeles & US & $1,360,462$ & $4.0$\\
Long Beach & US & $1,230,191$ & $4.0$\\
Vostochny & Russia & $931,440$ & $4.0$\\
Vancouver & Canada & $902,164$ & $4.0$\\
King Abdullah & Saudi Arabia & $848,920$ & $4.0$\\
Tacoma & US & $634,167$ & $4.0$\\
Seattle & US & $606,047$ & $4.0$\\
Prince Rupert & Canada & $120,980$ & $4.0$\\
Jiangyin & China & $30,156$ & $4.0$\\
Singapore & Singapore & $7,732,268$ & $3.8$\\
Port Said & Egypt & $1,798,944$ & $3.3$\\
Rotterdam & Netherlands & $4,343,461$ & $3.0$\\
Hamburg & Germany & $3,695,669$ & $3.0$\\
Port Kelang & Malaysia & $3,512,490$ & $3.0$\\
Felixstowe & UK & $2,378,804$ & $3.0$\\
Le Havre & France & $2,142,498$ & $3.0$\\
Bremerhaven & Germany & $1,905,953$ & $3.0$\\
Jeddah & Saudi Arabia & $1,874,875$ & $3.0$\\
Salalah & Oman & $1,673,569$ & $3.0$\\
Marsaxlokk & Malta & $1,462,921$ & $3.0$\\
Southampton & UK & $1,441,165$ & $3.0$\\
Khor Fakkan & UAE & $921,245$ & $3.0$\\
Zeebrugge & Belgium & $684,815$ & $3.0$\\
Sines & Portugal & $441,525$ & $3.0$\\
Wilhelmshaven & Germany & $415,918$ & $3.0$\\
Gdansk & Poland & $225,241$ & $3.0$\\
Kaliningrad & Russia & $182,586$ & $3.0$\\
Kwangyang & South Korea & $1,587,926$ & $2.9$\\
Aarhus & Denmark & $241,769$ & $2.9$\\
Trieste & Italy & $348,319$ & $2.8$\\
Rijeka & Croatia & $242,332$ & $2.8$\\ \hline
\end{tabular}
\end{minipage}
\end{tabular}
\end{table}
\end{document}